\documentclass[aps,preprintnumbers,floatfix,superscriptaddress]{revtex4}
\usepackage{graphicx,dcolumn,bm,epsfig,cancel,hyperref}
\usepackage{amsmath}
\usepackage{amssymb, times}

\def\fun#1#2{\lower3.6pt\vbox{\baselineskip0pt\lineskip.9pt
\ialign{$\mathsurround=0pt#1\hfil##\hfil$\crcr#2\crcr\sim\crcr}}}

\def\lsim{\mathrel{\rlap{\raise 2.5pt \hbox{$<$}}\lower 2.5pt\hbox{$\sim$}}}
\def\gsim{\mathrel{\rlap{\raise 2.5pt \hbox{$>$}}\lower 2.5pt\hbox{$\sim$}}}

\input epsf

\newcommand{\be}{\begin{equation}}
\newcommand{\ee}{\end{equation}}
\newcommand{\bea}{\begin{eqnarray}}
\newcommand{\eea}{\end{eqnarray}}

\newcommand{\comment}[1]{{}}

\usepackage{color}

\newcommand{\blue}{\textcolor{blue}}

\newcommand{\x}{\varphi}

\begin{document} 

\title{
Gravitational Waves, baryon asymmetry of the universe and electric dipole moment in the CP-violating NMSSM
}

%% %simple case: 2 authors, same institution
%% \author{A. Uthor}
%% \author{and A. Nother Author}
%% \affiliation{Institution,\\Address, Country}

% more complex case: 4 authors, 3 institutions, 2 footnotes
\author{Ligong Bian}
\email{lgbycl@cqu.edu.cn}
\affiliation{Department of Physics, Chongqing University, Chongqing 401331, China}
\affiliation{Department of Physics, Chung-Ang University, Seoul 06974, Korea}

\author{Huai-Ke Guo}
\email{ghk@itp.ac.cn}
\affiliation{CAS Key Laboratory of Theoretical Physics, Institute of Theoretical Physics, Chinese Academy of Sciences, Beijing 100190, China}
\author{Jing Shu}
\email{jshu@itp.ac.cn}
\affiliation{CAS Key Laboratory of Theoretical Physics, Institute of Theoretical Physics, Chinese Academy of Sciences, Beijing 100190, China}
\affiliation{
CAS Center for Excellence in Particle Physics, Beijing 100049, China
}
\affiliation{
School of Physical Sciences, University of Chinese Academy of Sciences, Beijing 100190, P. R. China
}
\affiliation{Center for High Energy Physics, Peking University, Beijing 100871, China}

% e-mail addresses: one for each author, in the same order as the authors

\begin{abstract}

In this work, we make the first study of electroweak baryogenesis (EWBG) based on the LHC data in the CP-violating next-to-minimal supersymmetric model (NMSSM) where a strongly first order electroweak phase transition (EWPT) is obtained in the general complex Higgs potential. %The CP-violating sources can and the CP-violation comes from the Higgs potential at tree level as well as the loop level. 
With representative benchmark points which pass the current LEP and LHC constraints, we demonstrate the structure of EWPT for those points and how a strongly first order EWPT is obtained in the complex NMSSM where the resulting gravitational wave production properties are found to be within the reaches of future space-based interferometers like BBO and Ultimate-DECIGO. We further calculate the generated baryon asymmetries where the CP violating sources are (1): higgsino-singlino dominated, (2): higgsino-gaugino dominated or (3): from both sources. It is shown that all three representing scenarios could evade the strong constraints set by various electric dipole moments (EDM) searches where cancellations among the EDM contributions occur at the tree level (higgsino-singlino dominated) or loop level (higgsino-gaugino dominated). The 125 GeV SM like Higgs can be either the second lightest neutral Higgs $H_2$ or the third lightest neutral Higgs $H_3$. Finally, we comment on the future direct and indirect probe of CPV in the Higgs sector from the collider and EDM experiments.

\end{abstract}
\maketitle
\preprint{}

\section{Introduction} 
The standard model (SM) of particle physics provides a satisfactory description of particle physics
phenomena over the past few decades together with the discovery of the Higgs boson~\cite{Aad:2012tfa, Chatrchyan:2012xdj}, a key element to the electroweak symmetry breaking mechanism.
However the theory is not yet perfect since the SM can not provide a dark matter candidate and furthermore can not explain the baryon asymmetry of the universe (BAU). The BAU is traditionally characterized by the baryon to entropy density ratio and has been measured to a high precision by Planck~\cite{Ade:2015xua},
\begin{eqnarray}
  Y_B = \frac{n_B}{s} = (8.61 \pm 0.09)\times 10^{-11}.
\end{eqnarray}
Among the various baryogenesis mechanisms, the electroweak baryogenesis (EWBG)\cite{Kuzmin:1985mm} is 
among the most theoretically well motivated and experimentally testable scenarios since it connects 
BAU generations to the details of the electroweak symmetry breaking (see~\cite{Morrissey:2012db} for a recent review). 
According to Sakharov~\cite{Sakharov:1967dj}, generation of a non-vanishing baryon asymmetry 
requires three ingredients in the particle physics of the early universe: baryon number violation,
C and CP-violation (CPV) and out of equilibrium conditions. 
In the framework of of EWBG, the non-equilibrium environment for 
baryon generation is provided through a first order electroweak phase transition(EWPT).
Even though the SM provides baryon number 
violation through the electroweak sphaleron process, it fails to provide a first order EWPT with
a $125\text{GeV}$ Higgs as well as a large enough CPV for sufficient baryon generation.
Physics scenarios beyond the SM, with the capability of providing a first order EWPT and with new sources of CPV, 
are therefore resorted to for a successful baryon number generation. 

Among the various new physics scenarios, the supersymmetric theory is among one of the most popular and intensively studied models due to its theoretical attractions that it can solve the gauge hierarchy problem, has the dark matter candidate, leads to strong-electroweak gauge unification and also can potentially explain the origin of baryon asymmetry in the universe. The minimal supersymmetric standard model(MSSM) accommodates two Higgs doublets, where the lightest of neutralinos can serve as a dark matter candidate and it can also provide new sources of CPV needed for baryon asymmetry. However the MSSM suffers from the $\mu$-problem~\cite{Kim:1983dt,Nir:1995bu,Cvetic:1995rj} and it is also hard to obtain a strongly first order EWPT (SFOEWPT) in this model~\blue{\cite{Carena:2012np,Curtin:2012aa,Cohen:2011ap,Cohen:2012zza}}. Both of these problems can be solved in a minimal way by extending the MSSM with an electroweak singlet chiral superfied. This is known in the literature as the next-to-minimal supersymmetric standard model (NMSSM)~\cite{Ellwanger:2009dp,Maniatis:2009re}. With this addition of the extra field content, the particle spectrum of NMSSM now accommodates two extra scalars giving then a total of five neutral scalars and a pair of charginos. This makes it quite easier to obtain a SFOEWPT in this model which has been studied in the CP conserving case~\cite{Pietroni:1992in,Davies:1996qn,Huber:2000mg,Funakubo:2005pu,Carena:2011jy,Balazs:2013cia,Kozaczuk:2014kva,Huang:2014ifa, Bi:2015qva,Huber:2015znp}. Besides, there can be CPV at tree level from the new interactions in contrast to the case in MSSM where CPV occurs at loop level. This new CPV in NMSSM is manifested in terms of a tree level mixing between the set of three CP-even scalars and the set of two CP-odd scalars, similar to what happens in the two Higgs doublet model~\cite{Shu:2013uua, Inoue:2014nva,King:2015oxa,Bian:2016zba}. 
%\blue{The EDMs measurements constraints on tree level and loop level CPV phases has been investigated independently in Ref.~\cite{King:2015oxa}.}
With the presence of these two kinds of CPV sources, there appears the chance in the parameter space that their CPV effects cancel~\cite{Bian:2014zka} in contributions to the electric dipole moments (EDM) of electron, neutrons and atoms, etc,. and therefore can evade the stringent constraints from null search results of the EDMs. Moreover, even though the studies of BAU in the context of MSSM and in the NMSSM have been performed in the literatures~\cite{Cheung:2012pg,Kozaczuk:2013fga}, a joint analysis with these two different origins of CPV is still lacking. We therefore make an updated study of the EDM and BAU phenomenologies  with these two different CPV sources taken into account. Furthermore, with the addition of the gauge singlets as well as the inclusion of the CPV in the Higgs potential, the effective potential is now of relatively high dimension and this can potentially leads to a rich structure of phase transition patterns. In particular, the phase transition does not necessary proceed through one step to the electroweak vacuum and generally need multiple steps~\cite{Funakubo:2005pu} for this to happen which therefore deserves a detailed scrutiny. In addition, after the discovery of the gravitational waves from merging black holes detected by LIGO~\cite{Abbott:2016blz}, there has been increasing interests in the discussions of the gravitational waves from the EWPT in various new physics models (see Ref.~\cite{Cai:2017cbj} for a recent review). Therefore it deserves a similar study in the NMSSM to augment the analysis of the baryogenesis and EDMs.

The rest of this paper is organized as follows. We first introduce our conventions for the complex 
NMSSM in Sec.~\ref{sec:model}. We explore the phase structures of this model in Sec.~\ref{sec:EWPT} where the impact of CPV on the EWPT is studied with representative benchmark parameter space points. We then present the properties of gravitational wave signals generated during the EWPT for these benchmark points in the following Sec.~\ref{sec:gw}. We perform a joint analysis of the baryon asymmetry in the framework of 
EWBG and the constraints from null searches of EDM in Sec.~\ref{sec:bauedm} and briefly comment on the collider probes of CPV in Sec.~\ref{sec:DP} after which we give the summary in Sec.~\ref{sec:Summary}.

%\vspace{3cm}

\section{The CP-violating Higgs sector \label{sec:model}}
We consider the $Z_3$-invariant complex NMSSM in which the Higgs superpotential and the soft supersymmetry 
breaking terms are given respectively by~\cite{Ellwanger:2009dp} 
\begin{align}
W_\mathrm{Higgs} &=  
  \lambda \widehat{S}\,\widehat{H}_u \cdot \widehat{H}_d + \frac{\kappa}{3}\widehat{S}^3 , \\
-{\Delta \cal L}_\mathrm{soft}&\supset \lambda A_{\lambda}H_{u}\cdot H_{d}S + \frac{1}{3}\kappa A_{\kappa}S^{3}+\mathrm{h.c.}~,
\end{align}
with here the fields with a hat being chiral superfields and those without a hat being the corresponding 
scalar components. Here the limit $\kappa \rightarrow 0$ corresponds to the Pecci-Quinn limit.
Collecting the F-terms from above superpotential $W_{\text{Higgs}}$, the Higgs interactions from
$\Delta \mathcal{L}_{\text{soft}}$ and the D-terms from the Higgs gauge interactions, one obtains 
the tree-level Higgs potential:
\begin{align}
\label{V0}
V_0=V_F+V_D+V_{\rm soft},
\end{align}
and each of above contributions is given explicitly by,
\begin{align}
&V_F=|\lambda|^2|S|^2(H_d^\dagger H_d+H_u^\dagger H_u)
        +|\lambda H_u \cdot H_d+\kappa S^2|^2,\nonumber \\
&V_D=\frac{g_2^2+g_1^2}{8}(H_d^\dagger H_d-H_u^\dagger H_u)^2
        +\frac{g_2^2}{2}(H_d^\dagger H_u)(H_u^\dagger H_d),\nonumber \\
&V_{\rm soft}=m_{H_d}^2 H_d^\dagger H_d+m_{H_u}^2 H_u^\dagger H_u
        +m_S^2|S|^2 +(\lambda A_{\lambda}S H_u \cdot H_d
        +\frac{1}{3}\kappa A_\kappa S^3+{\rm h.c.}).
\end{align}
We take here the parameters $\lambda$, $A_{\lambda}$, $\kappa$ and $A_{\kappa}$ to be generically complex 
as opposed to the CP-conserving case where these parameters are real. Thus these complex 
parameters provide additional sources of CPV aside from those originally appearing in the 
MSSM case.

After the electroweak symmetry breaking, two further phases $\theta_u^0$ and $\theta_S^0$ appear in the 
expansion of the Higgs fields about the vacuum expectation values(VEVs) of the neutral components of $H_u$
, $H_d$ and $S$ which are defined with the following convention~\cite{Funakubo:2005pu},
\begin{eqnarray}
H_d&=&
\hphantom{e^{i\theta}}\,
\hspace{0.2cm}
\left(
\begin{array}{c}
\frac{1}{\sqrt{2}}\,(v_d+h_d+ia_d) \\
\phi_d^-
\end{array}
%\right), \quad
\right), \nonumber \\[1mm]
%H_u=
H_u&=&
e^{i\theta_u^0}\,\left(
\begin{array}{c}
\phi_u^+\\
\frac{1}{\sqrt{2}}\,(v_u+h_u+ia_u)
\end{array}
\right), \nonumber \\[1mm]
S\ \ &=&\frac{e^{i\theta_S^0}}{\sqrt{2}}\,(v_S+h_S+ia_S)\,.
\label{eq:higgsparam}
\end{eqnarray}
Here $v_u$ and $v_d$ are the VEVs of $H_u$ and $H_d$ respectively, $h_u$, $h_d$, $h_S$ are the CP-even 
scalars, $a_u$, $a_d$, $a_S$ are the CP-odd scalars and we have removed the phase of $H_d$ by the gauge 
transformations. Then, the effective higgsino 
mixing parameter in NMSSM is given by $\mu=|\lambda| v_s e^{i \phi_{\mu}}/\sqrt{2}$ with $\phi_{\mu}=\theta_s^0+\phi_\lambda$.

\comment{
\textcolor{blue}{I will add the other three minimization conditions and also all mass matrix elements.}
}

For translating into physical parameters, 
three of the minimization conditions about the VEVs allow us to replace the soft mass parameters 
$m_{H_u}^2$, $m_{H_d}^2$ and $m_{S}^2$ by $v_u$, $v_d$ and $v_S$ while the remaining three conditions
yield relations among the complex parameters indicating that the CP phases are not all independent.
Considering this, we follow Ref.~\cite{Funakubo:2005pu} and introduce the following notations to make our discussions 
independent of the phase conventions,
\begin{eqnarray}
\label{eq:cpvrp}
\mathcal{R} \hspace{0.15cm} &=& |\lambda| |\kappa|\, \cos(\phi^\prime_\lambda-\phi^\prime_\kappa)\,,
\hspace{1.5cm}
\mathcal{I} \hspace{0.15cm} = |\lambda| |\kappa|\, \sin(\phi^\prime_\lambda-\phi^\prime_\kappa)\,,
\nonumber \\
R_\lambda &=& \frac{|\lambda| |A_\lambda|}{\sqrt{2}}\,
	\cos(\phi^\prime_\lambda+\phi_{A_\lambda})\,,
\hspace{1cm}
I_\lambda = \frac{|\lambda| |A_\lambda|}{\sqrt{2}}\,
	\sin(\phi^\prime_\lambda+\phi_{A_\lambda})\,,
\nonumber \\
R_\kappa &=& \frac{|\kappa| |A_\kappa|}{\sqrt{2}}\,
	\cos(\phi^\prime_\kappa+\phi_{A_\kappa})\,,
\hspace{1cm}
I_\kappa = \frac{|\kappa| |A_\kappa|}{\sqrt{2}}\,
	\sin(\phi^\prime_\kappa+\phi_{A_\kappa})\,,
%R_q &=& \frac{|\lambda| |A_q|}{\sqrt{2}}\,
%	\cos(\phi^\prime_\lambda+\phi_{A_q})\,,
%I_q = \frac{|\lambda| |A_q}{\sqrt{2}}\,
%	\sin(\phi^\prime_\lambda+\phi_{A_q})\,,
%\nonumber \\
\label{eq:defRI}
\end{eqnarray}
with here $\phi^\prime_\lambda \equiv \phi_\lambda+\theta_u^0+ \theta_S^0 \,,
\phi^\prime_\kappa \equiv \phi_\kappa+3\theta_S^0\,$ and 
$\phi_{\lambda}^{\prime}$, $\phi_{A_{\lambda}}$, $\phi_{\kappa}^{\prime}$, $\phi_{A_{\kappa}}$ are 
the phases of $\lambda$, $A_{\lambda}$, $\kappa$, $A_{\kappa}$ respectively. Here $I_{\lambda}$ 
and $I_{\kappa}$ denote the CP phases of the first and second terms in the soft SUSY breaking Lagrangian
and due to the minimization conditions, they are related to the single CPV source $\mathcal{I}$ by
\begin{align}
\label{eq:tdcpv}
I_\lambda &= - \frac{1}{2}\,\mathcal{I}\, v_S\,, \hspace{1.5cm}
I_\kappa = -\frac{3}{2}\,\mathcal{I}\, \frac{v_dv_u}{v_S} .
\end{align}
So there is only one physical CP phase at the tree level: $\phi_\lambda^\prime - \phi_\kappa^\prime$ while 
$\phi_\lambda^\prime+A_\lambda$ and $\phi_\kappa^\prime+A_\kappa$ can be determined from this 
up to a two-fold ambiguity.
However above conditions face modificatioins by loop corrections and this will be discussed in the 
one-loop effective potential in the following.

With the presence of the Higgs sector CPV corresponding to 
$\phi^\prime_\lambda - \phi^\prime_\kappa \neq 0$, 
the neutral Higgs bosons need not carry any definite CP parities
already at the tree level and their mixing can be described by
an real orthogonal $5\times 5$ matrix $O$ as
\begin{equation}
\label{basis}
%\Phi =
\left( \phi^0_d\,, \phi^0_u\,, \phi^0_S\,, a\,, a_S \right)^T \ = \
O_{\alpha i}
\left( H_1\,, H_2\,, H_3\,, H_4\,, H_5 \right)^T ,
\end{equation}
with here $H_{1(5)}$ defined as the lightest (heaviest) Higgs mass eigenstate,
$a \equiv a_d \sin\beta+a_u\cos\beta$ and the massless Goldstone boson 
$G_0 \equiv a_u \sin\beta-a_d\cos\beta$ decouples from above scalars in the quadratic terms.
This mixing constitutes the most significant difference compared with the MSSM case,
in which the CPV can only be induced at one-loop level, 
mostly through loop corrections to the Higgs self-energy, 
from the phase of the soft SUSY breaking trilinear couplings $A_u, A_d, A_e$ of the up-type, 
down-type and charged lepton-type sfermions, as well as the soft SUSY breaking mass parameters 
of the gauginos $M_1, M_2$ and $M_3$. 
To capture the main features of EDMs and EWBG, we study the scenarios with 
the tree-level CPV phase coming from $\phi_\kappa$ and the loop-level induced CPV phase being $\phi_\mu^\prime\equiv \phi_\mu-\phi_{M_2}$. To describe the EDMs properties clear, the three phases of $\phi_{\kappa,\mu,M_2}$ will be used as input parameters in the numerical analysis of Sec.~\ref{sec:bauedm}. 

\section{Electroweak Phase Transition \label{sec:EWPT}}
The phase structure of the model is governed by the finite temperature effective potential $V_{\text{eff}}$, 
composed of the following contributions,
\begin{align}
\label{Veff}
    &V_{\rm eff} = V_{\rm Tree} + V_{\rm CW} + V_{\rm CT} + V_{\rm T} .
\end{align}
Firstly, the tree level effective potential $V_{\rm Tree}$ is given by Eq.~(\ref{V0}),
\begin{align}
&&  V_{\rm Tree} = \frac{1}{2}m_{H_d}^2 \x_d^2 + \frac{1}{2} m_{H_u}^2 \x_u^2 
+ \frac{1}{2}m_S^2\x_S^2 + \frac{g_2^2 + g_1^2}{32}(\x_u^2 - \x_d^2)^2 
+ \frac{|\lambda|^2}{4}(\x_d^2 + \x_u^2)\x_S^2  \nonumber \\
&& \hspace{1cm} + \left|-\frac{\lambda}{2}e^{i\theta_u}\x_d\x_u + \frac{\kappa}{2}e^{2i\theta_S}\x_S^2\right|^2 
+ \left(-\frac{\lambda A_\lambda}{2\sqrt{2}}e^{i(\theta_u+\theta_S)}\x_d\x_u\x_S + \frac{\kappa A_\kappa}{6\sqrt{2}}e^{i3\theta_S}\x_S^3 +\rm{h.c.}\right) ,
\end{align}
where $\x_d, \x_u, \x_S, \theta_u$ and $\theta_S$ are the background fields of the Higgs scalars. 
For convenience of the calculations of EWPT, we construct the 5-dimensional (5D) order parameters of EWPT~\cite{Funakubo:2005pu},
\begin{align}
\label{orderpara}
  \x_i = (\x_d, \x_u\cos\Delta\theta_u, \x_u\sin\Delta\theta_u, \x_S\cos\Delta\theta_S, \x_S\sin\Delta\theta_S) ,
\end{align}
with here $\Delta\theta_u = \theta_u - \theta_u^0\,$ and $\Delta\theta_S = \theta_S - \theta_S^0\,$. 
Using these parameters and with the definitions in Eq.~(\ref{eq:defRI}), the the tree-level contribution 
can be rewritten as
\begin{align}
\label{Vtree}
    V_{\rm Tree} &= \frac{1}{2}m_{H_d}^2\x_1^2 + \frac{1}{2}m_{H_u}^2(\x_2^2 + \x_3^2) 
    + \frac{1}{2}m_{S}^2(\x_4^2 + \x_5^2) \notag\\
    &+ \frac{g_2^2+g_1^2}{32}(\x_1^2 - \x_2^2 - \x_3^2)^2 + \frac{|\kappa|^2}{4}(\x_4^2 + \x_5^2)^2 \notag\\
    &+ \frac{|\lambda|^2}{4}\left[(\x_1^2 + \x_2^2 + \x_3^2)(\x_4^2 + \x_5^2) + \x_1^2(\x_2^2 + \x_3^2) \right] \notag\\
    &+ R_\lambda\x_1(\x_3\x_5-\x_2\x_4) + I_\lambda\x_1(\x_2\x_5+\x_3\x_4)
    + \frac{1}{3}R_\kappa\x_4(\x_4^2 - 3\x_5^2) + \frac{1}{3}I_\kappa\x_5(\x_5^2 - 3\x_4^2) \notag\\
    &- \frac{1}{2}\x_1\mathcal{R}\left(\x_2(\x_4^2 - \x_5^2) + 2 \x_3\x_4\x_5\right)
    +\frac{1}{2}\x_1\mathcal{I}\left(\x_3(\x_4^2 - \x_5^2) - 2 \x_2\x_4\x_5\right) .
\end{align} 

The second part in the effective potential is the one-loop Coleman-Weinberg potential 
$V_{\rm CW}$~\cite{Coleman:1973jx}, 
given by
\begin{align}
\label{Vcw}
& V_{\rm CW} = \sum_i \frac{(-)^{2s_i}n_i}{64\pi^2}m_i^4(\vec{\x})\left[\ln\frac{m_i^2(\vec{\x})}{Q^2}
 -c_i\right]  ,
\end{align}
where $i$ runs through all particles in the model, with degrees of freedom $n_i$, field-dependent mass 
$m_i(\vec{\x})$ and spin $s_i$. 
The above result is given in the Landau gauge with renormalization scale $Q$ and
the constant $c_i$ is $5/6$ for gauge bosons and $3/2$ for the others. 
With above Coleman-Weinberg potential included, the tadpole conditions are modified by these 
one-loop corrections. Hence, the counterterms $V_{\rm CT}$ are introduced to preserve the tree-level 
relations for VEVs~\cite{Cline:2011mm}, i.e. 
\begin{align}
\label{ctship}
    &\left.{\partial V_{\rm CT}\over \partial \x_l}\right|_{\rm VEVs} = 
-\left.{\partial V_{\rm CW}\over \partial \x_l}\right|_{\rm VEVs} ,
\end{align}
where it is sufficient to include the following terms in $V_{\text{CT}}$:
\begin{align}
\label{Vct}
    V_{\rm CT} &= \frac{1}{2}\delta m_{H_d}^2\x_1^2 + \frac{1}{2}\delta m_{H_u}^2(\x_2^2 + \x_3^2) 
    + \frac{1}{2}\delta m_{S}^2(\x_4^2 + \x_5^2)
     + \delta I_\lambda\x_1(\x_2\x_5+\x_3\x_4) .
     %\blue{+ \frac{1}{3}\delta I_\kappa\x_5(\x_5^2 - 3\x_4^2)}
\end{align}
%\blue{Is it true that $\delta I_\kappa=0$ ?}

The last piece in the effective potential is the thermal corrections $V_{\rm T}$ 
at the one-loop level~\cite{Quiros:1999jp},
\begin{equation}\label{VT}
  V_{\rm T} = \frac{T^4}{2\pi^2} \sum_{i} (-1)^{2 s_i} n_i \int_0^\infty dx\;x^2 \ln(1\mp e^{-\sqrt{x^2+m_i^2/T^2}}).
\end{equation}
Note these lowest order thermal corrections need to be improved by daisy 
resummations~\cite{Parwani:1991gq,Gross:1980br} at high temperatures 
when the perturbative expansion of the potential fails. This in practice can be done by insertions of 
thermal mass corrections, $m_i^2\rightarrow m_i^2 + \delta m_i^2$, in Eq.~(\ref{VT})~\cite{Cline:2011mm}.
Here, we note that the Goldstone contributions may leads to a negative $m_i^2$ which has been treated as in Ref.~\cite{Bernon:2017jgv}. 

With the finite temperature effective potential given above, the phase structure of the model can be readily obtained by searching for its minima at each step as temperature drops. The basic picture of the phase evolution is as follows. At sufficiently high temperature, the universe
sits at the origin of the 5D field space where the electroweak symmetry is manifest.
 \begin{table}[!ht]
 \begin{center}
 \scalebox{1}{
  \begin{tabular}{c|c|c|c|c|c|c|c|c} \hline
  $\lambda$ &$\mu$ (GeV) & $A_\lambda$ (GeV) & $\kappa$ & $A_\kappa$ (GeV) & tan $\beta$&$A_{t}(\text{GeV})$ \\ \hline
   0.6 &180 & 335.0 & 0.13 & -99.18 &1.5&1402  \\ \hline
    $A_b(\text{GeV})$&$A_\tau(\text{GeV})$&$M_2(\text{GeV})$&$M_1(\text{GeV})$&$m_{h_2}$ (\text{GeV}) & PTS(CPC)&PTS(CPV)  \\ \hline
 1539&1502&-601&-102&125 &1.043 &1.039 \\ \hline
  \end{tabular}}
 \end{center}
 \caption{The CP conserving NMSSM benchmark point with a small $\tan\beta$, the CPV scenario is obtained by tuning on CPV phases $\phi_{M_2}=0.08$ and $\phi_\kappa=0.08$. The PTS represents the phase transition strength given by $\phi_{\text{EW}}(T_C)/T_C$. \label{nubauedm}}
\end{table}
As temperature drops, other minima will develop in this space and the universe will transit to 
the lower minimum. Depending on the presence or absence of a barrier between these two adjacent minima, this phase transition is classified as first order or second order respectively. The phase transition occurs through nucleations of the bubble within which the lower minimum is developed~\cite{Coleman:1977py,Linde:1980tt,Linde:1981zj}. The bubbles expand, 
collide and coalesce with each other leaving eventually the universe in the vacuum of the lower minimum. During this process, the temperature at which these two minima are degenerate is the critical temperature $T_C$. For baryon asymmetry to be generated within the non-equilibrium 
phase boundaries, the phase transition that is connected to the electroweak symmetry breaking needs to be strongly first order to sufficiently quench the weak sphaleron process inside the bubble and thus to preserve the generated baryons. This generally translates into the widely adopted 
criteria $\phi_{\text{EW}}(T_C)/T_C \gtrsim 1$~\cite{Shaposhnikov:1986jp,Shaposhnikov:1987tw,Cline:2006ts}
where $\phi_{\text{EW}} \equiv \sqrt{\varphi_u^2+\varphi_d^2}$
\footnote{Both the $\phi_{\text{EW}}(T_C)$ and $T_C$ involve gauge dependence issues~\cite{Patel:2011th}. One should keep in mind that we analyze the EWPT in Landau gauge when interpreting our results.}.
\begin{figure}[t]
\centering
\includegraphics[width=0.53\textwidth]{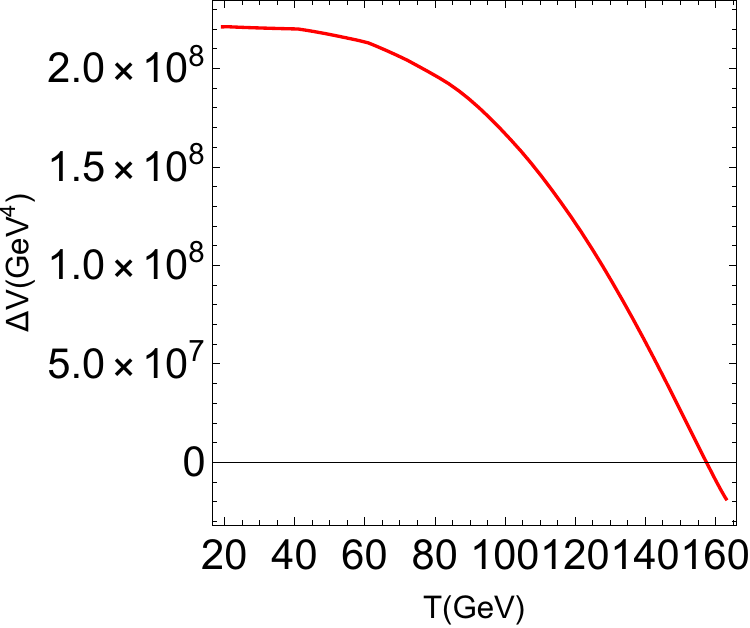}
\includegraphics[width=0.46\textwidth]{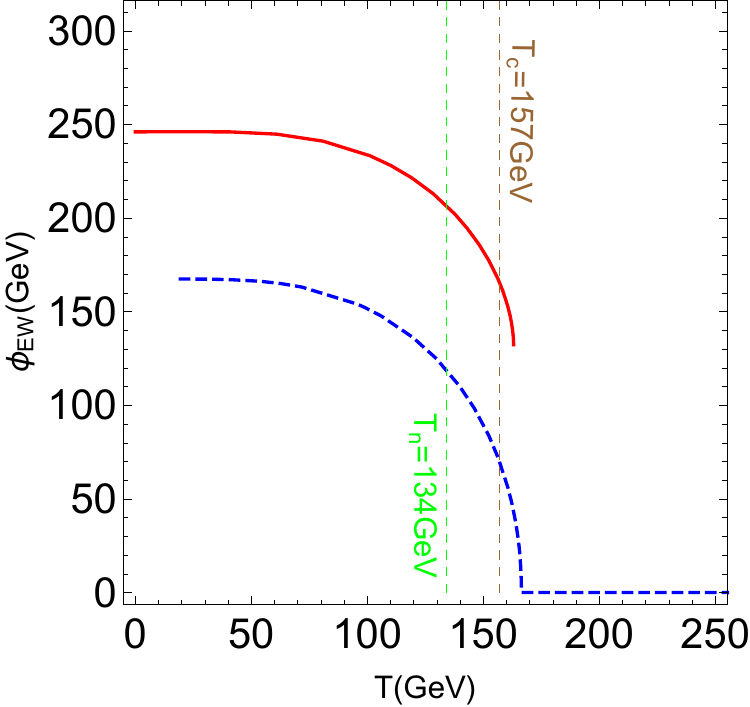}
\\

\includegraphics[width=0.53\textwidth]{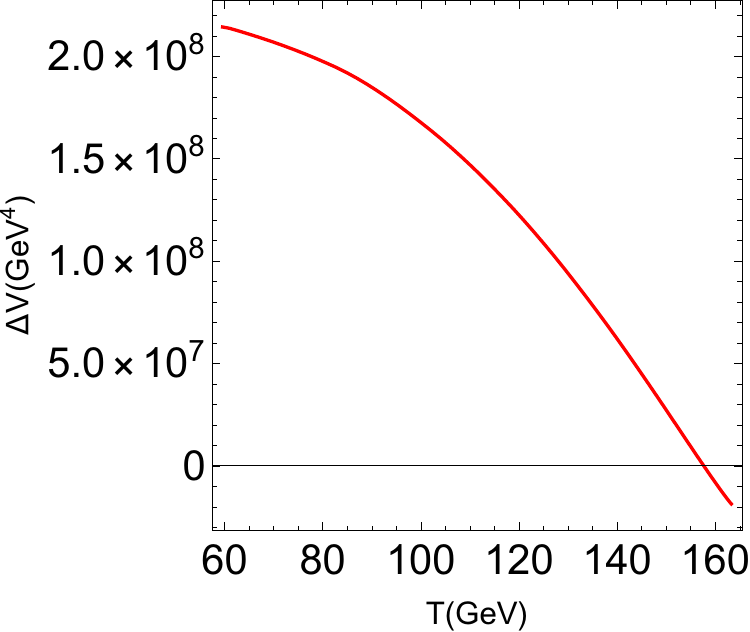}
\includegraphics[width=0.46\textwidth]{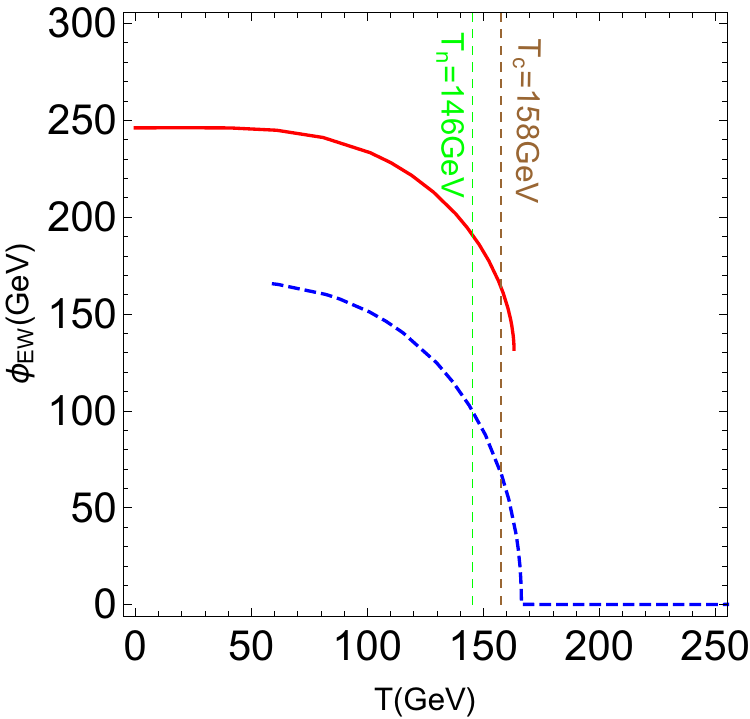}
\caption{
\label{fig:phase}
The evolution of $\Delta V$(left) and $\phi_{\text{EW}}$ (right)
as temperature drops from right to left for the CP-conserving small $\tan\beta$ benchmark point of 
Table~\ref{nubauedm} (top). The bottom plots are for the CP-violating case obtained as described in 
the caption of Table.~\ref{nubauedm}.
For the right plots, also plotted are the critical temperature $T_C$ in brown and the nucleation temperature $T_n$ in green. 
}
\end{figure}

Due to the addition of the extra scalar singlet and the inclusion of CPV in the Higgs potential, 
the field space is of relatively high dimension and thus the phase transition history can be of quite 
rich structures~\cite{Funakubo:2005pu} and typically have a two step 
EWPT~\cite{Cheung:2013dca,Patel:2012pi,Jiang:2015cwa,Shu:2006mm,Inoue:2015pza,Patel:2013zla,Chala:2016ykx,Vaskonen:2016yiu,Huber:2015znp,Chao:2017vrq}~\footnote{Notice that those models can be probed through the resonance di-Higgs searches~\cite{Kang:2013rj,No:2013wsa,Huang:2017jws}.}. 
Due to the relatively large parameter space of this model, we seek here only representative benchmark 
points to illustrate the phase structures while a more comprehensive analysis of the model parameter 
space is deferred to future works. In order to show the effect of CPV to the EWPT, here we present 
and compare two representative benchmark points corresponding to the CP-conserving and CP-violating cases 
respectively, with the numerical results here obtained using {\bf CosmoTransitions}~\cite{Wainwright:2011kj}. 
The resulting phase structures for the 
CP-conserving case is shown in the top panel of Fig.~\ref{fig:phase} corresponding to the benchmark point in 
Table.~\ref{nubauedm}. The CP-violating case, which is obtained as described in the caption 
Table.~\ref{nubauedm}, is shown in the bottom panel.
For both cases, the right plots show the evolution of the electroweak background field amplitude
$\phi_{\text{EW}}(T)$ as temperature drops 
from right to left. Here blue lines denotes the high temperature phases which are all observed to 
leave the origin continuously, thus signifying its second order nature. As temperature decreases, 
the phases corresponding to the red lines appear and denote the minima which eventually evolve to 
the electroweak minima. During the temperature ranges where these two phases overlap, the critical 
temperatures are shown with brown dashed vertical lines and is $157~ \text{GeV}$ for the CP-conserving 
case while a slightly larger value of $158 ~\text{GeV}$ is found when the CPV is turned on.
So the inclusion of CPV tends to drive the onset of EWPT earlier while the duration of the high phase is 
seen to be shortened.
Also shown in these plots are the nucleation temperature to be defined in the following section. For both cases, the left 
plots show the evolution of the difference of $V_{\text{eff}}$ at the high and low phases during the 
overlapped temperature ranges, where the critical temperatures correspond to $\Delta V=0$. 
As was observed in these plots, the EWPT for these two benchmark points proceeds through two steps with 
the first step being second order and the subsequent one being first order with its strength being
$\phi_{\text{EW}}(T_C)/T_C=1.039$, consistent with the SFOEWPT criterion~\footnote{
For the second step EWPT, the weak sphaleron process outside the electroweak bubble is suppressed 
compared with that in the symmetric phase since electroweak symmetry is already broken outside.  A
benchmark point with a better EWPT pattern could generally be found from a dedicated scan 
over the NMSSM parameter space which however is beyond the scope of this work.
}.

\section{Gravitational Waves\label{sec:gw}}
During the first order EWPT, there can be gravitational waves generated, mainly coming from three 
processes: bubble collisions, sound waves in the plasma and Magnetohydrodynamic 
turbulence~\cite{Caprini:2015zlo} (see~\cite{Cai:2017cbj} for a recent review). The total energy density of 
the resulting gravitational waves is approximately a sum of these three contributions,
\begin{eqnarray}
  \Omega_{\text{GW}}h^2 \simeq \Omega_{\text{col}} h^2 + \Omega_{\text{sw}} h^2 + \Omega_{\text{turb}} h^2,
\end{eqnarray}
where we adhere to the Hubble constant definition $H = 100 h \text{km}\ \text{s}^{-1}\ \text{Mpc}^{-1}$. 
The energy spectrums depends on the electroweak bubble profiles during the EWPT which corresponds to the 
bounce solutions minimizing the action,
\begin{eqnarray}
S_3(T)=\int d^3 x \left[ \frac{1}{2} (\vec{\nabla} \phi_b)^2 + V(\phi_b,T)  \right],
\end{eqnarray}
leading to the equation of motion for solving the bubble profile $\phi_b$ ,
\begin{eqnarray}
\frac{d^2 \phi_b}{d r^2} + \frac{2}{r} \frac{d \phi_b}{d r} - \frac{\partial V(\phi_b, T)}{\partial \phi_b} = 0.
\end{eqnarray}
In addition, for a bubble to be effectively growing, triumphing the force from surface tension, 
the following condition needs to be met~\cite{Apreda:2001us}
\begin{eqnarray}
\int_0^{t_{\ast}} \frac{\Gamma}{H^3} dt = \int_{T_{\ast}}^{\infty} \frac{d T}{T}\left(\frac{90}{8\pi^3 g}\right)^2 \left(\frac{M_{Pl}}{T}\right)^4 e^{-S_3(T)/T} \sim 1\ ,
\end{eqnarray}
which serves as the definition of the nucleation temperature $T_n$.
This then translates into finding the nucleation temperature $T_n$ such that  
$S_3(T_{\ast})/T_{\ast}|_{T_{\ast} =T_n} \approx 140$~\cite{Apreda:2001us} is satisfied.
From the bubble profiles and the nucleation temperature, the following two key parameters $\alpha$ and 
$\beta$, relevant for gravitational wave calculations, can be obtained 
\begin{eqnarray}
  \alpha = \frac{30 \Delta \rho}{\pi^2 g_{\ast}T_{\ast}^4}  \bigg|_{T_{\ast} \approx T_n}\ ,
\hspace{2cm}
\frac{\beta}{H_n} = T_{\ast} \frac{d(S_3/T)}{d T} \bigg|_{T_{\ast} = T_n} ,
\end{eqnarray}
where in the evaluation of $\alpha$, $\Delta \rho$ is the difference of energy density between the 
false and true vacua~\cite{Kozaczuk:2014kva},
\begin{eqnarray}
\rho(v(T_n), T_n) = - V(v(T), T) + T \frac{d}{T} V(v(T), T) \bigg|_{T_{\ast} \approx T_n} ,
\end{eqnarray}
and $H_n$ is the Hubble constant evaluated at the nucleation temperature $T_n$.
With these parameters solved, one can obtain the energy spectrum of the gravitational waves.

Firstly for the gravitational waves from the bubble collision, its contribution can be estimated using the envelop approximation~\cite{Kosowsky:1991ua,Kosowsky:1992rz,Kosowsky:1992vn} and results in the following spectrum~\cite{Huber:2008hg},
\begin{eqnarray}
  \Omega_{\text{col}} h^2 = 1.67\times 10^{-5} \left(\frac{H_{\ast}}{\beta}\right)^2
  \left(\frac{\kappa \alpha}{1+\alpha}\right)^2 \left( \frac{100}{g_{\ast}} \right)^{1/3}
  \left( \frac{0.11 v_w^3}{0.42 + v_w^2} \right) 
  \frac{3.8(f/f_{\text{env}})^{2.8}}{1+2.8(f/f_{\text{env}})^{3.8}} \ , 
\end{eqnarray}
with here $v_w$ being the bubble wall velocity and $\kappa$ characterizing the fraction of latent heat
deposited in a thin shell, both of which are functions of the previously defined parameter $\alpha$~\cite{Kamionkowski:1993fg},
\begin{eqnarray}
v_w \simeq \frac{1/\sqrt{3} + \sqrt{\alpha^2+2\alpha/3} }{1+\alpha},  \nonumber \quad \quad
\kappa \simeq  \frac{0.715\alpha + \frac{4}{27} \sqrt{3\alpha/2}}{1+0.715\alpha} .
\end{eqnarray}
Moreover $f_{\text{env}}$ is the peak frequency at 
present time and is approximately given by
\begin{eqnarray}
  f_{\text{env}} = 16.5\times 10^{-6} \left(\frac{f_{\ast}}{\beta}\right) 
  \left(\frac{\beta}{H_{\ast}}\right) \left(\frac{T_{\ast}}{100\text{GeV}}\right)
  \left(\frac{g_{\ast}}{100}\right)^{1/6}
  \text{Hz} .
\end{eqnarray}
Secondly, for the contribution from sound waves, it is given by 
\begin{eqnarray}
  \Omega_{\text{sw}} h^2 = 2.65\times 10^{-6} \left(\frac{H_{\ast}}{\beta}\right)
  \left(\frac{\kappa_v \alpha}{1+\alpha}\right)^{2} \left(\frac{100}{g_{\ast}}\right)^{1/3} v_w 
  \left(\frac{f}{f_{\text{sw}}}\right)^3 \left( \frac{7}{4+3(f/f_{\text{sw}})^2} \right)^{7/2} .
\end{eqnarray}
Here $\kappa_v$ is the fraction of latent heat transformed into the bulk motion of the fluid and
is approximately given by $\kappa_v \approx \alpha(0.73+0.083\sqrt{\alpha} + \alpha)^{-1}$ in the 
case of $v_w\approx 1$ while for the peak frequency $f_{sw}$, we have 
\begin{eqnarray}
  f_{\text{sw}} = 1.9\times 10^{-5} \frac{1}{v_w} 
  \left(\frac{\beta}{H_{\ast}}\right) \left(\frac{T_{\ast}}{100\text{GeV}}\right) 
  \left(\frac{g_{\ast}}{100}\right)^{1/6} \text{Hz} .
\end{eqnarray}
Finally the MHD turbulence contributes
\begin{eqnarray}
  \Omega_{\text{turb}} h^2 = 3.35\times 10^{-4} \left(\frac{H_{\ast}}{\beta}\right)
  \left(\frac{\kappa_{\text{turb}} \alpha}{1+\alpha}\right)^{3/2} 
  \left(\frac{100}{g_{\ast}}\right)^{1/3} v_w 
  \frac{(f/f_{\text{turb}})^3}{[1+(f/f_{\text{turb}})]^{11/3} (1+8\pi f/h_{\ast})},\nonumber \\
\end{eqnarray}
with here $\kappa_{\text{turb}} \approx 0.1 \kappa_v$ and in this case, the peak frequency $f_{\text{turb}}$ is
\begin{eqnarray}
  f_{\text{turb}} = 2.7\times 10^{-5} \frac{1}{v_w} 
  \left(\frac{\beta}{H_{\ast}}\right) \left(\frac{T_{\ast}}{100\text{GeV}}\right) 
  \left(\frac{g_{\ast}}{100}\right)^{1/6} \text{Hz} .
\end{eqnarray}

\begin{figure}[t]
  \centering
 \includegraphics[width=0.49\textwidth]{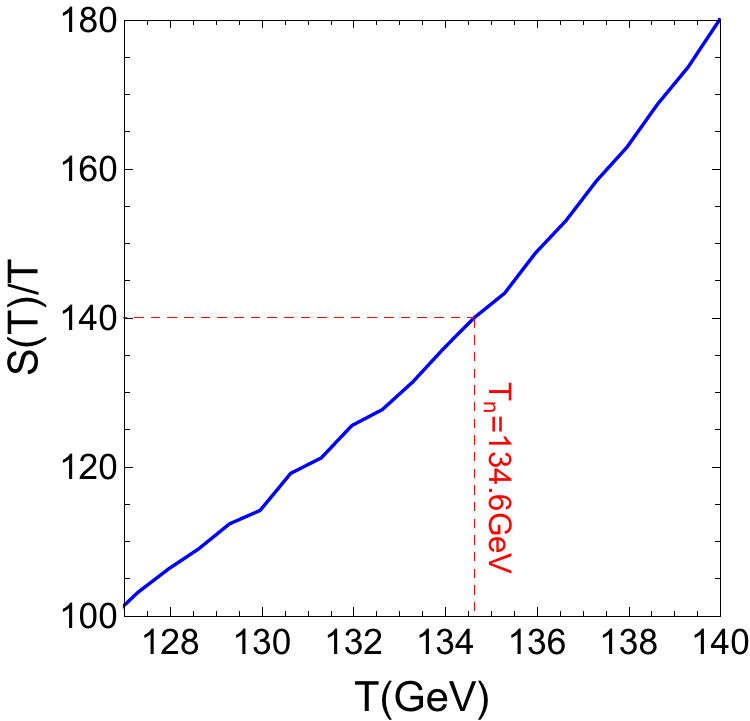}
 \quad 
  \includegraphics[width=0.47\textwidth]{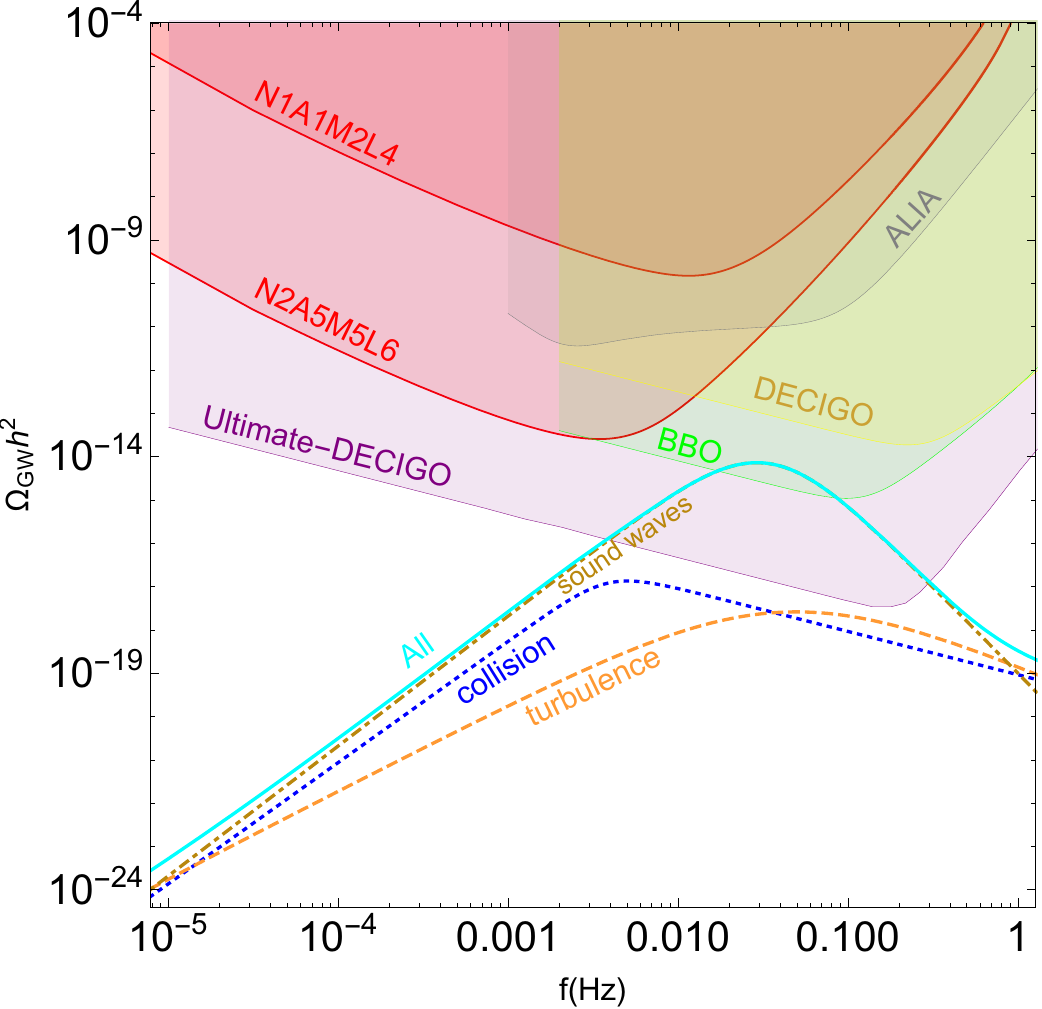}
\caption{\label{fig:gw}
Left panel: the behavior of $S_3(T)/T$ in the neighborhood of the nucleation temperature $T_n = 134~\text{GeV}$ as the temperature drops from right to left for the benchmark point of Table.~\ref{nubauedm}. Right panel: The energy spectrum of gravitational waves as function of the frequency generated during the strongly first order EWPT from three sources: turbulence(yellow dashed line), bubble collision(blue dotted line) and sound waves(brown dotdashed line) for the benchmark point of Table.~\ref{nubauedm}. The total energy density is shown as a cyan line and is almost indistinguishable from the brown line since it is dominated by the sound waves 
contribution in this scenario. The color shaded regions at the top are  experimentally sensitive regions for LISA (two configurations with notation NiAjMkLl), ALIA (gray), BBO (green), DECIGO (yellow) and Ultimate-DECIGO (purple).
}
\end{figure}

\begin{figure}[t]
  \centering
  \includegraphics[width=0.55\textwidth]{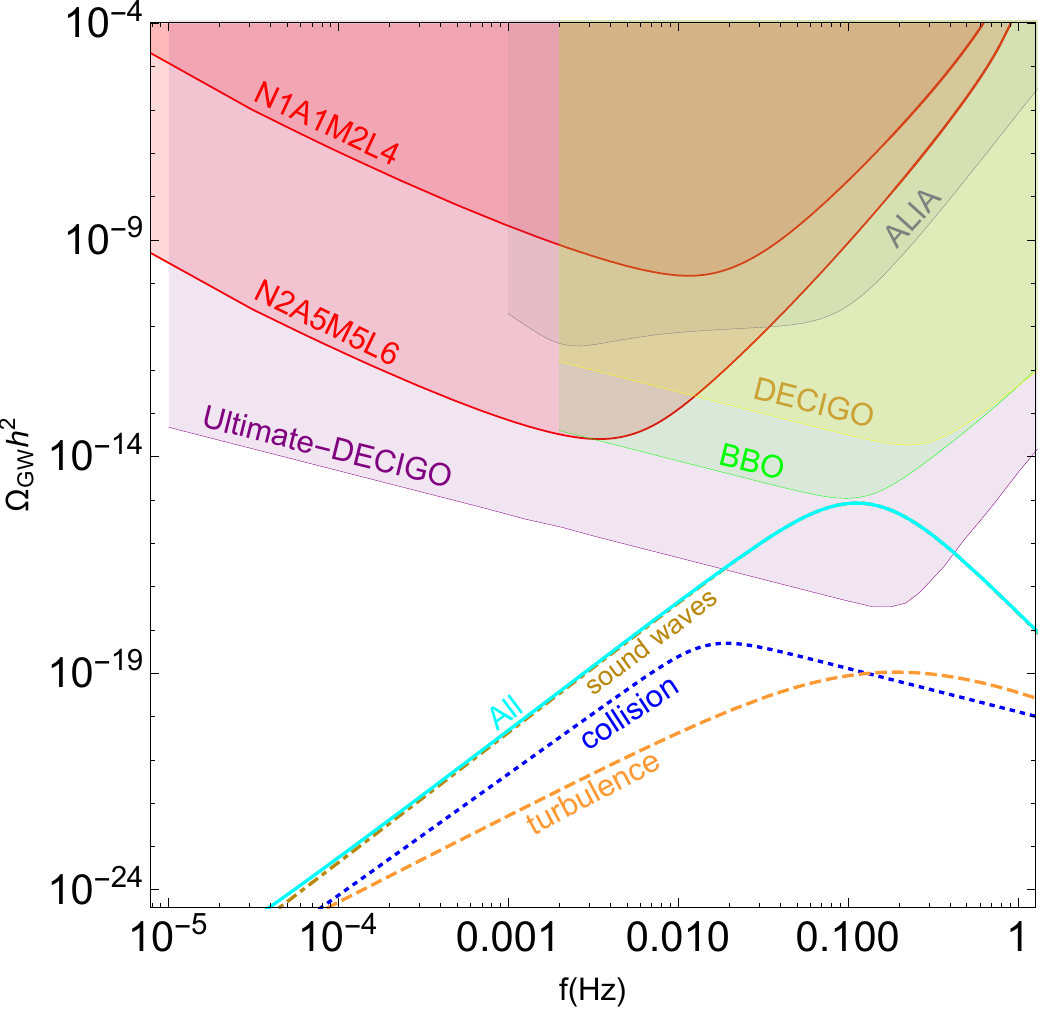}
\caption{\label{fig:gw-cpv}
The gravitational wave energy spectrums for the small $\tan\beta=1.5$ case NMSSM with CPV corresponding to
Table~\ref{nubauedm} and the bottom plots in Fig.~\ref{fig:phase}.
The plotting conventions are the same as that in the right plot of Fig.~\ref{fig:gw}.
}
\end{figure}

After the discovery of the SM-like Higgs at LHC, the gravitational waves produced by bubble collision in the 
CP-conserving NMSSM has been studied in Ref.~\cite{Kozaczuk:2014kva}. The gravitational waves produced by 
the bubble collision and sound waves has been studied later~\cite{Huber:2015znp}. 
We explore the prospects of GW signals for the CP-conserving and CP-violating benchmarks presented in 
Table.~\ref{nubauedm}. Specifically we use the package {\bf CosmoTransitions}~\cite{Wainwright:2011kj} to solve 
the bounce solutions and to find the nucleation temperatures $T_n$ with which we calcualte the 
parameters $\alpha$, $\beta$ and finally the gravitational wave spectrums. 
In this process, 
we observe that the benchmarks both satisfy the nucleation criterior $S_3(T)/T < 140$ below 
certain nucleation temperatures $T_n$. The calculated nucleation temperatures for these two 
cases are shown in the respective plots in Fig.~\ref{fig:phase} with green
dashed vertical lines. 
We further show the resulting gravitational wave spectrums for these two cases: Fig.~\ref{fig:gw} for the 
CP-conserving case and Fig.~\ref{fig:gw-cpv} for the CP-violating case. In the left plot of 
Fig.~\ref{fig:gw}, we also show the profile of $S_3(T)/T$ in the neighborhood of its nucleation 
temperature $T_n = 134.6~\text{GeV}$ for illustration.
In these gravitational wave spectrum plots, the three individual contributions from bubble collision,
sound waves and turbulence are ploted using blue dotted, brown dotdashed and yellow dashed lines with their
sum denoted by a solid cyan line. In each of these cases, the sound wave contribution dominates and is 
almost indistinguishable with the cyan line. The color shaded regions in these plots are the experimental
sensitivity regions for several proposed space based interferometers: LISA~\cite{Audley:2017drz} with two design 
configurations in notation NiAjMkLl~\cite{Caprini:2015zlo,Klein:2015hvg}, BBO, 
DECIGO (Ultimate-DECIGO)~\cite{Kudoh:2005as} and ALIA~\cite{Gong:2014mca}
For the CP-conserving case corresponding to Fig.~\ref{fig:gw}, the gravitational wave energy spectrum 
falls within the sensitivies of BBO and Ultimate-DECIGO while unreachable by the others. Turning on CPV as 
plotted in Fig.~\ref{fig:gw-cpv}, the magnitude of the energy spectrum decreases a little bit and can 
barely be detected by BBO. 

We note again that, these benchmark points presented here only constitute a tiny fraction of the whole 
parameter space of the NMSSM and therefore there might be other parameter space points which can give 
significantly enhanced GW spectrum. Exploration of this vast parameter space however requires a significantly improved calculations of the bounce solutions, a task which we will defer to future works.

\section{Anatomy of BAU and EDM \label{sec:bauedm}}
In this section, we study the explanation of BAU in the framework of EWBG under the constraints of 
EDMs measurements. Typically, we focus on the CPV phases coming from tree level and loop level together.

\subsection{The Electroweak Baryognesis in the CPV NMSSM\label{sec:BAU}} 
With the tree level CP-phases taken into account, there are in general three different CPV source terms in the quantum transport equations governing the dynamics of the particle densities in the framework of the closed-time-path-formalism (CTP) (see Ref.~\cite{Lee:2004we}
for pedagogical discussions). Under assumptions of chemical equilibrium of Yukawa interactions, strong sphaleron and gaugino interactions, the set of coupled transport equations can be reduced to a single equation of 
${H}$~\cite{Lee:2004we,Chung:2009cb,Chung:2009qs},
\begin{eqnarray}
  v_w {H}^{\prime}(\bar{z}) - \bar{D} {H}^{\prime}(\bar{z}) 
  + \bar{\Gamma} {H}(\bar{z}) - \bar{S} =0,
\label{eq:H}
\end{eqnarray}
where a one dimensional picture of the expanding bubble wall assumption is chosen as usually 
adopted in the literature to simplify the calculations, $v_w$ is the bubble wall velocity and 
$\bar{z}$ is the spatial coordinate perpendicular to the wall in the
frame where the wall is at rest with negative values of $\bar{z}$ corresponding
to the symmetric electroweak phase (bubble exterior). It should be note that the wall velocity for the gravitional wave and baryogenesis studies should be different due to the the fist one require a relatively larger $v_w$ in comparison with 
the latter one. The $v_w$ for the calculation of gravational wave and baryogenesis can be different when taking into account the hydrodynamics of the bubble~\cite{No:2011fi,Caprini:2011uz}, the consistent calculation can be performed with the one for gravational wave being the speed of the wall and the one for the calculation of baryogenesis being the relative velocity between the wall and the plasma.

In the above equation, the quantities $\bar{D}$, $\bar{\Gamma}$ and $\bar{S}$ are the 
effective diffusion constant, effective relaxation rate and effective source term for ${H}(\bar{z})$ 
respectively and the explicit formulae are~\cite{Chung:2009cb,Chung:2009qs},
\begin{subequations}\label{eq:difr}
\begin{align}
\bar D&=
\frac{D_H+D_Q(\kappa_T-\kappa_B)+D_L \kappa_L}{1+\kappa_T-\kappa_B+
 \kappa_L}\,, \\
\label{eq:gammabar}
\bar \Gamma&=
\frac{\Gamma_h+\Gamma_{mt}+\Gamma_{mb}+\Gamma_{m\tau}}
{k_H(1+\kappa_T-\kappa_B+\kappa_L)}\,,\\
\bar S&=\frac{S^{\cancel{CP}}_{\widetilde H}+S^{\cancel{CP}}_{\widetilde t}
-S^{\cancel{CP}}_{\widetilde b}+ S^{\cancel{CP}}_{\widetilde \tau}
}{1+\kappa_T-\kappa_B+ \kappa_L}\,.
\end{align}
\end{subequations}
Here the Higgs rate $\Gamma_h$ is 
\begin{eqnarray}
\Gamma_h = \frac{6}{T^2}\,\left(
\Gamma_{\widetilde{H}^\pm\widetilde{W}^\pm} +
\Gamma_{\widetilde{H}^0\widetilde{W}^0} +
\Gamma_{\widetilde{H}^0\widetilde{B}^0} +
\Gamma_{\widetilde{H}^0\widetilde{S}}
\right) ,
\end{eqnarray}
and the relaxation rates for the stop, sbottom, and stau cases 
could be computed following 
Ref.~\cite{Lee:2004we, Chung:2008aya, Chung:2009cb, Cirigliano:2006wh, Chung:2009qs}. 
The contributions from the Higgs sector CPV are summarized in the quantity 
$S^{\cancel{CP}}_{\widetilde H}$ given by
\begin{equation}
S^{\cancel{CP}}_{\widetilde H}=
S_{\widetilde H^\pm}^{\cancel{CP}}(x)
+S_{\widetilde{H}^0}^{\cancel{CP}}(x)
+S_{\widetilde{S}\widetilde{H}^0}^{\cancel{CP}}  ,
\label{eq:CPVSH}
\end{equation}
Here the first term $S^{\cancel{CP}}_{\widetilde H^{\pm}}$ is the gaugino-higgsino driven CP-violating 
source term and can be computed in the vev insertion approximation from a tree-level wino $\widetilde{W}^{\pm}$ mediation. The result is given by~\cite{Lee:2004we}
\begin{eqnarray}
\label{eq:chargino3}
&& S_{\widetilde H^\pm}^{\cancel{CP}}(x) =\frac{g_2^2}{\pi^2}v(x)^2\dot\beta(x)
M_2\mu\sin\phi_{\mu}^\prime \times\int_0^\infty\frac{dk\,k^2}{\omega_{\widetilde H}\omega_{\widetilde W}}\nonumber\\
&&
\times \text{Im}
\biggl[ \frac{n_F(\mathcal{E}_{\widetilde W}) - n_F(\mathcal{E}_{\widetilde H}^*)}{(\mathcal{E}_{\widetilde W} - \mathcal{E}_{\widetilde H}^*)^2}
- \frac{n_F(\mathcal{E}_{\widetilde W}) + n_F(\mathcal{E}_{\widetilde H})}{(\mathcal{E}_{\widetilde W} + \mathcal{E}_{\widetilde H})^2}\biggr]\,, 
\label{eq:CPVSCH1}
\end{eqnarray}
where $\dot{\beta}(x)=d\beta/{dt}$ and can be approximated by $\dot{\beta} \approx v_{w} \Delta \beta/L_w$ with here
$v_w$ being the bubble wall velocity, $L_w$ the wall width and $\Delta \beta$ the difference of the 
value $\beta(x)$ outside and inside the bubble. 
So above source term grows linearly with $\Delta\beta$ and our choice is 
optimal. This dependence of $\Delta\beta$ is to be regarded as 
a theoretical uncertainty and the precise determination of $\Delta\beta$ in the NMSSM is however 
beyond the scope of this work. For the more precise calculation, the profiles of the bubble wall
and the profile-dependent masses and widths %of the higgsino and singlino 
should be taken into account when solving the quantum transport equations during the EWPT.

The second term $S_{\widetilde H^0}^{\cancel{CP} }(x)$ in Eq.~(\ref{eq:CPVSH}) denotes the contribution 
from the neutral higgsino
mixing terms and the corresponding result for $\widetilde W$ and $\widetilde B$ intermediate states can be obtained from Eq.~(\ref{eq:CPVSCH1}) by the replacements: $g_2\rightarrow g_2/\sqrt{2}$ and
$g_2\rightarrow g_1/\sqrt{2}$, $\omega_{\tilde{W}}\rightarrow\omega_{\tilde{B}}$, 
$\Gamma_{\widetilde{W}}\rightarrow\Gamma_{\widetilde{B}}$, 
 
\begin{eqnarray}
\label{eq:chargino4}
 S_{\widetilde H^0}^{\cancel{CP}}(x) &&=\frac{g_2^2}{2\pi^2}v(x)^2\dot\beta(x)
M_2\mu\sin\phi_{\mu}^\prime \times\int_0^\infty\frac{dk\,k^2}{\omega_{\widetilde H}\omega_{\widetilde W}}\nonumber\\
&&
\times \text{Im}
\biggl[ \frac{n_F(\mathcal{E}_{\widetilde W}) - n_F(\mathcal{E}_{\widetilde H}^*)}{(\mathcal{E}_{\widetilde W} - \mathcal{E}_{\widetilde H}^*)^2}
-\frac{n_F(\mathcal{E}_{\widetilde W}) + n_F(\mathcal{E}_{\widetilde H})}{(\mathcal{E}_{\widetilde W} + \mathcal{E}_{\widetilde H})^2}\biggr]\,\nonumber\\
&&+\frac{g_1^2}{2\pi^2}v(x)^2\dot\beta(x)
M_1\mu\sin\phi_{\mu} \times\int_0^\infty\frac{dk\,k^2}{\omega_{\widetilde H}\omega_{\widetilde B}}\nonumber\\
&&
\times \text{Im}
\biggl[ \frac{n_F(\mathcal{E}_{\widetilde B}) - n_F(\mathcal{E}_{\widetilde H}^*)}{(\mathcal{E}_{\widetilde B} - \mathcal{E}_{\widetilde H}^*)^2}
- \frac{n_F(\mathcal{E}_{\widetilde B}) + n_F(\mathcal{E}_{\widetilde H})}{(\mathcal{E}_{\widetilde B} + \mathcal{E}_{\widetilde H})^2}\biggr]\,.
\label{eq:CPVSCH2}
\end{eqnarray}

%For the thermal width of the singlino, we are taking
%$\Gamma_{\widetilde{S}}=0.03\,T$ considering the large
%coupling $|\lambda|=0.81$. 
%We find $Y_B$ is affected by the amount of about (25-35) \%
%as $\Gamma_{\widetilde{S}}$ varies between $0.003\,T$ and $0.03\,T$.
Since these above two source terms depend on $\sin\phi_{\mu}^\prime$ and therefore they vanish 
when the phase $\phi_{\mu}^\prime$ is set to zero.
For further notations about the quantities appearing here, we refer the readers to~\cite{Lee:2004we}. 
We note that in the NMSSM, these gaugino-higgsino sources driven EWBG have been 
studied in~\cite{Kozaczuk:2013fga}, wherein tree level CP phases are all imposed to be zero~\cite{Chung:2009cb}.
The third term in Eq.~(\ref{eq:CPVSH}) is the higgsino-singlino driven CPV source and has been studied 
in Ref.~\cite{Cheung:2012pg},
\begin{eqnarray}
\hspace{1cm} S_{\widetilde{S}\widetilde{H}^0}^{\cancel{CP}} 
= 
-2|\lambda|^2|M_{\widetilde{S}}||\mu|v^2
	\dot{\beta}\sin(\phi_\lambda-\phi_\kappa)\,
	\mathcal{I}_{\widetilde{S}\widetilde{H}^0}^f \;, 
	\label{eq:shbau}
\end{eqnarray}
where 
\begin{eqnarray}
\hspace{1cm}
|M_{\tilde S} (T)| =
\left[ 2 |\kappa|^2 v_S^2 +
\frac{|\lambda|^2+2|\kappa|^2}{8}\,T^2 
\right]^{1/2} ,
\end{eqnarray}
including the singlino thermal mass term. 
Here we have assumed that there is no spontaneous CPV and the functional form
of the Fermionic source function $\mathcal{I}_{\tilde{S}\tilde{H}^0}^f$ can be found in Ref.~\cite{Cheung:2012pg}.
It is obvious from Eq.~(\ref{eq:shbau}) that $S_{\widetilde{S}\widetilde{H}^0}^{\cancel{CP}}$ vanishes 
when the phase combination vanishes, that is, when $\sin(\phi_\lambda - \phi_\kappa)=0$.

Aside from the Higgs sector CP-violating source terms in Eq.~(\ref{eq:CPVSH}), we also included the source terms from sfermion sector:
\begin{equation} 
\begin{aligned}
S_{\tilde{f}}^{\cancel{CP}}(x)=
\frac{N_C^f y_f^2}{2\pi^2}\operatorname{Im}(\mu A_f)v^2(x)\dot{\beta}(x) 
\int_0^{\infty} \frac{dk k^2}{\omega_R \omega_L}   
\operatorname{Im}
\left[\frac{n_B(\mathcal{E}^*_R)-n_B(\mathcal{E}_L)}{(\mathcal{E}_L-\mathcal{E}_R^*)^2}
+\frac{n_B(\mathcal{E}_R)+n_B(\mathcal{E}_L)}{(\mathcal{E}_L+\mathcal{E}_R)^2} \right] .
 \end{aligned}
\end{equation} 
Since the source terms are negligibly small for $\bar{z}<-L_w/2$, and that the
relaxation terms have the approximate form $\bar\Gamma(\bar{z})=\theta(\bar{z})\bar \Gamma$,
the solution of $H$ in the symmetric phase accepts the following solution from Eq.~(\ref{eq:H}),
\begin{equation}
H={\cal A}{\rm e}^{v_w \bar{z}/{\bar D}}\,.
\end{equation}
with the prefactor given by
\begin{eqnarray}
\label{eq:amplitude}
&& {\cal A}=\int\limits_0^\infty dy\,\bar S(y)
\frac{{\rm e}^{-\gamma_+ y}}{\bar D\gamma_+} 
+\!\!\!\int\limits_{-L_w/2}^0 dy\,\bar S(y)\left(
\frac{\gamma_-}{v_w\gamma_+}+\frac{{\rm e}^{-v_wy/{\bar D}}}{v_w}
\right)\,,
\end{eqnarray}
where the usually encountered factor $\gamma_{\pm}$ is,
\begin{equation}
\label{eq:gammapm}
\gamma_\pm=\frac{1}{2\bar D}\left(
v_w\pm\sqrt{v_w^2+4\bar\Gamma \bar D}
\right)\,.
\end{equation}

With the profile for $H$ solved under previous assumptions, the profiles for the other particle densities can be obtained. In particular the left-handed Fermionic charge density, which serves as the source for the weak sphaleron process for generating the baryons, can be obtained,
\begin{eqnarray}
\label{n:left}
&& \quad n_{\rm left}= H \times 
\left(
\frac{k_q}{k_H}\frac{k_B-k_T}{k_B+k_Q+k_T}
+\vartheta_L \frac{k_\ell}{k_H}\frac{k_R D_R}{k_L D_L + k_R D_R}
\right)\; .
\end{eqnarray}
The left handed density gets converted into a baryon density $n_B$ through weak sphaleron transitions.
The obtained baryon number density in the electroweak broken phase is a constant given by~\cite{Huet:1995sh},
\begin{equation}
\label{nB:sphal}
n_{B}=-3\frac{\Gamma_{\rm ws}}{v_{\rm w}}
\int\limits_{-\infty}^0 dz\; n_{\rm left}(z)
{\rm e}^{\frac{15}{4} \frac{\Gamma_{\rm ws}}{v_{\rm w}}z}\,.
\end{equation}
In above formulation, we have used the approximation that 
due to the much smaller weak sphaleron rate $\Gamma_{\rm ws}$ and thus
the rates for both the creation of $n_\mathrm{left}$ and its diffusion ahead of the bubble wall,  
above Eq.~(\ref{nB:sphal}) is usually decoupled from the set of diffusion equations.
\subsection{CPV Phases v.s. EDM in CPV NMSSM}

In the MSSM, the CPV could only occur at loop level. Assuming
mass hierarchy between the first two generations of sfermions, sleptons and the third generation,
the EDMs could be dominated by the Barr-Zee diagrams.    
Details on this case can be found in our previous work~\cite{Bian:2014zka}.
%It should be mentioned that, EWBG could be dominated by the CP-violating source arising
%from higgsino-wino mixing, which lives in the served region of EDMs measurements.
When the phase combination $\phi^\prime_\lambda - \phi^\prime_\kappa= 0$ and the phases of  $A_u, A_d, A_e,M_1,M_2,M_3$ are nonzero, the CPV in NMSSM occurs at loop level just like the CPV MSSM case, and the constraints from the EDMs measurements are supposed to be smaller than the case that CPV occurs at tree level. In this situation, EWBG could be dominated by the higgsino-wino mixing CP-violating source, i.e., Eq.~(\ref{eq:CPVSCH1}, \ref{eq:CPVSCH2}).
For the case of phase combination $\phi^\prime_\lambda - \phi^\prime_\kappa\neq0$ 
and the phases of $A_u, A_d, A_e,M_1,M_2,M_3$ being zero. The CPV of NMSSM could occur at tree level and the EWBG is driven by the higgsino-siglino CPV source, i.e., Eq.~(\ref{eq:shbau}). 
It should be mentioned that, the cancellations of the theoretical predictions of electron EDM (eEDM) between the CP phase at tree level and the one at loop level may help evade a lot of the parameter spaces from the stringent ACME2013 constraint~\cite{Baron:2013eja}, as will be explored with the CP phases of the Higgs sectors ($\phi_\kappa$, $\phi_{\mu}$) and chargino sector ($\phi_{M_2}$).
$\phi_{M_2}$ characterizes the couplings of $H-\chi^{\pm}-\chi^\pm$ and thus determines
the magnitude of W-loop contribution to Barr-Zee diagram of eEDM. It should be mentioned that, 
the phase $\phi_{M_2}$ also plays an important role in the coupling of $W^{\pm}-\chi^{\pm}-\chi^0$, 
thus a larger magnitude of $\phi_{M_2}$ might make a larger contribution to eEDM through 
the Barr-Zee diagrams with $\gamma W^\pm W^\mp$ coupling~\cite{Ellis:2010xm}.

%\begin{figure}[t]
%\centerline{
%\includegraphics[width=0.9\columnwidth]{figbau.pdf}\vspace{-0.3cm}}
%\caption{Values of $\Delta\theta$ and $\tan\beta$ consistent with the observed baryon asymmetry of the universe. We take the constraints from Fig.~\ref{Fig:GF} which point to $\alpha=\beta-\pi/2$, and $\Delta\theta=\alpha_b$. The blue star point is the benchmark point from the numerical studies in Ref.~\cite{Fromme:2006cm} that $\Delta \theta = \xi =0.2$ for $m_{h_1}= 125$ GeV and $m_{h_2}$ = 400 GeV. 
%\vspace{-0.4cm}}
%\label{bau}
%\end{figure}

\subsection{Numerical analysis of EWBG and EDM\label{sec:bauEDM}}

In this section, we calculate the Higgs spectrum using {\bf NMSSMCALC}~\cite{Baglio:2013iia}, implement the EDM analysis with the similar approach as Ref.~\cite{Bian:2014zka}, and show the combined results of EWBG and EDMs for two 
typical scenarios after imposing the constraints from {\bf HiggsBounds}~\cite{Bechtle:2013wla}: the CPV NMSSM with a small $\tan\beta$ and the CPV NMSSM with a moderate $\tan\beta$ (the semi-PQ limit case with $\kappa\approx 0$). In both scenarios, we vary the imaginary parts of $M_2$, $\mu$ and $\kappa$ to study the different CP phases of $\phi_{M_2}$, $\phi_\mu$ and $\phi_\kappa$, thereby taking into account both loop and tree level CPV
effects. In both cases the neutron EDM~\cite{Baker:2006ts} and Mercury EDM experimental constraints~\cite{Graner:2016ses} can be satisfied
for parameter spaces shown in the figures~\footnote{If we decrease the current neutron or Mercury EDM upper bound by an order of magnitude, in both cases all parameter spaces shown in the paper are ruled out, which indicates that those region will be probed in the near future by EDM experiments.}. 
  
%\begin{figure}[!ht]
 %\includegraphics[angle=0,scale=0.3]{edmbau.pdf}\\
   %\caption{\label{fig:}
 % The combined results of eEDM and BAU for PQ limit cases with varying $\lambda$ and $\phi_\kappa$, with the green region excluded by the current eEDM experiments, and the orange region favored by BAU experiments. }
 % \vspace{-0.3cm}
%\end{figure}

For the eEDM calculations in above two benchmark scenarios, both cases embrace the same 
property, that is, the top, W- and chargino- loop Barr-Zee diagrams dominate the eEDM contributions and the cancelation among these makes the magnitude of eEDM falling into the allowed regions set by the experiment ACME 2013.  
We perform the numerical analysis of eEDM in the scenarios that $m_{H_2}$ or $m_{H_3}$ is the SM-like Higgs
respectively. 
For the scenario with $m_{H_2}$ being the SM-like Higgs, we explore the case with a small $\tan\beta=1.5$.  
At last, the PQ-limit case with a relatively moderate $\tan\beta$ and with $\lambda\gg\kappa$ is 
studied in the parameter space of $\phi_{M_2}$ and $\phi_\kappa$.

\begin{figure*}[t]
 \includegraphics[width=0.3\textwidth]{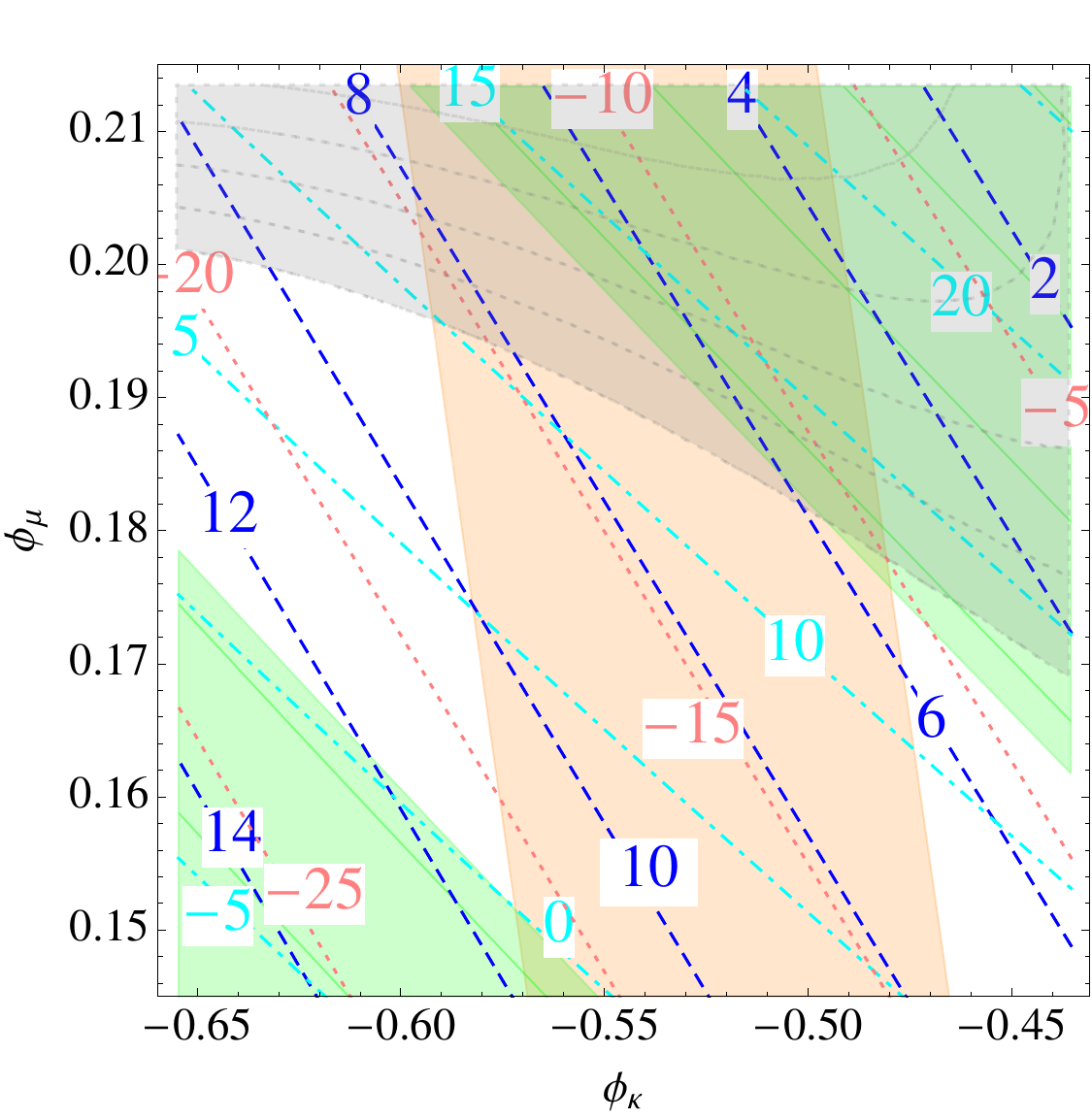}
 \quad
  \includegraphics[width=0.3\textwidth]{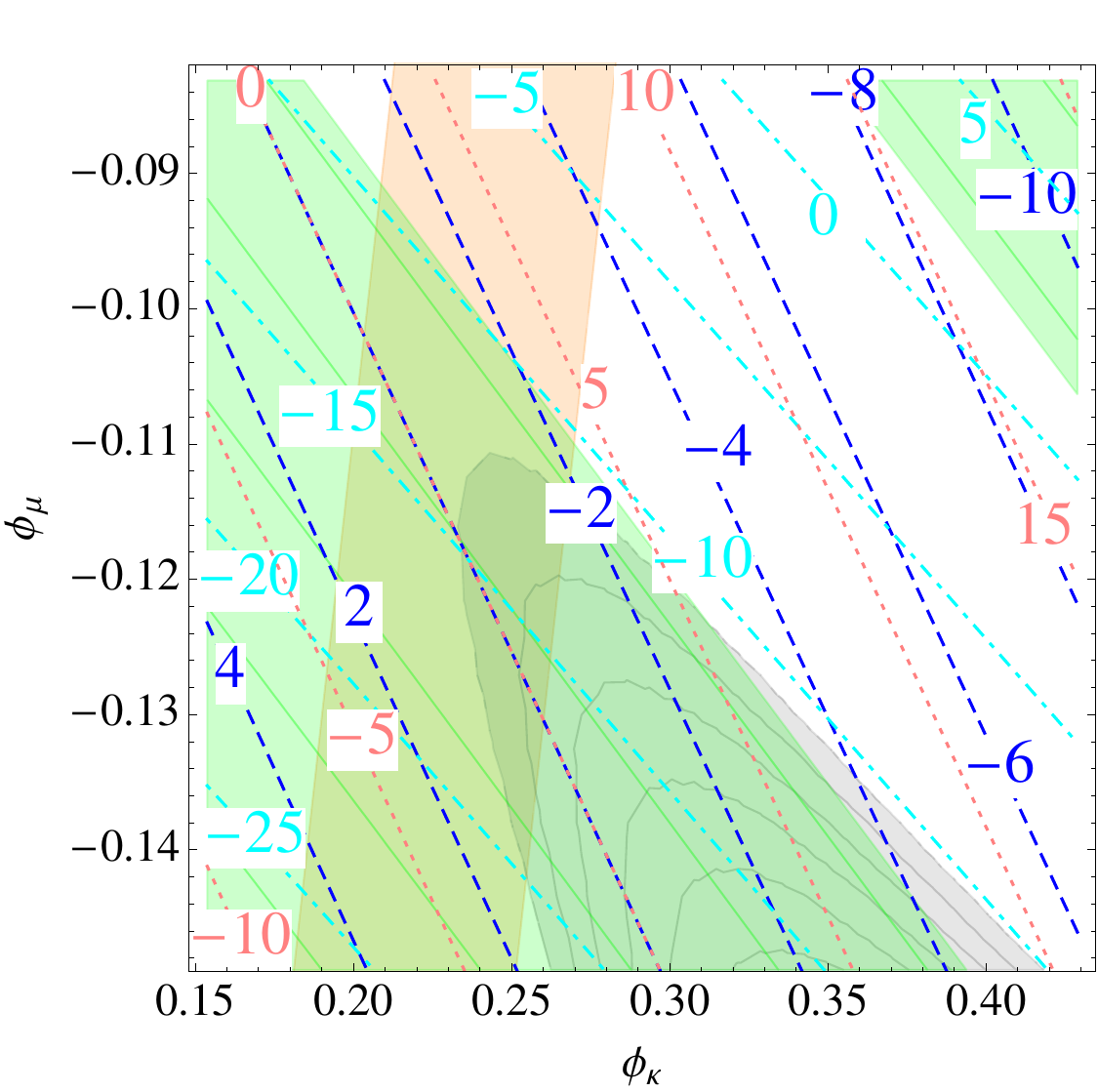}
  \quad
  \includegraphics[width=0.3\textwidth]{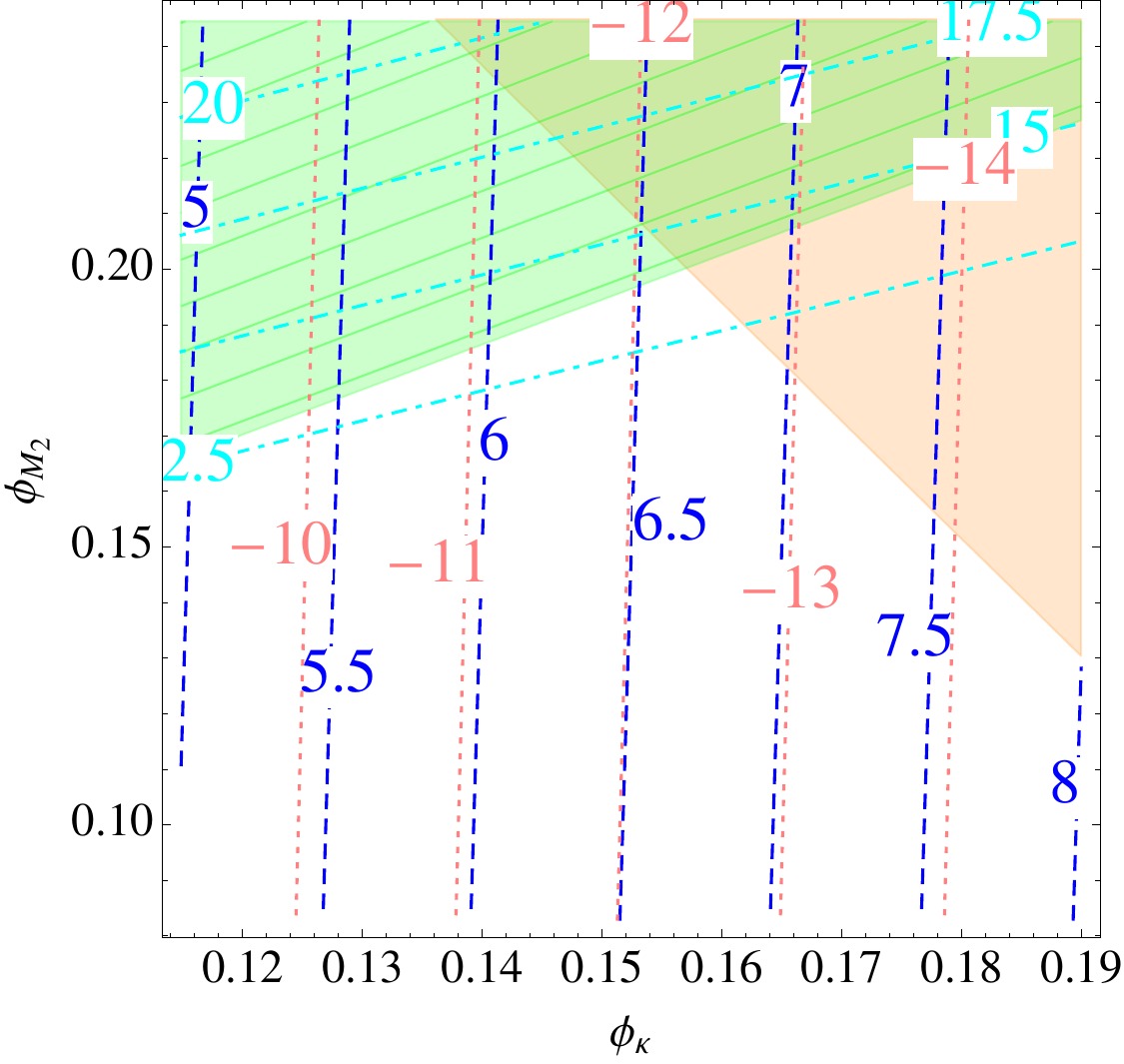}
   \caption{\label{fig:H2nm}
%%Fig.5.
The combined plots of the eEDM and BAU in the plane $(\phi_{\kappa}, \phi_{\mu}(\phi_{M_2}))$ with $\phi_{M_2}(\phi_{\mu})=0$ for the scenario that $H_2$ is the SM-like Higgs in the CPV NMSSM.
In both plots the green regions are excluded by the current eEDM experiments, the gray regions are excluded by requiring $m_{H_2}$ to be close to $125~\text{GeV}$ while the orange regions are favored by the BAU results. For the contours, the blue, cyan and pink dashed lines represent the magnitudes of top quark, charginos and $W$ boson loop contributions of the $\gamma H$ Barr-Zee diagrams of eEDM. The other parameters are chosen 
as in Table.~\ref{nubauedm}.  }
\end{figure*}
%
%\item The case of $H_2$ being SM-like Higgs in the natural NMSSM 
\subsubsection{$H_2$ as the SM-like Higgs with a small $\tan\beta$}

In this section, we study the scenario of $H_2$ being the SM-like Higgs and work with a small $\tan\beta=1.5$. We investigate CPV physics in 
the parameter space of
$(\phi_{\kappa}, \phi_{M_2})$ and 
$(\phi_{\kappa}, \phi_{\mu})$. Different from the case of Ref.~\cite{Bian:2014zka}, here the eEDM is mostly dominated by the $H_2$ and $H_3$
mediated Barr-Zee diagrams contributions.
In this case, the pseudo-scalar $a_s$ constitutes the main proportion of $H_3$, and the $H_2$ is dominated by $h_u$ with $a_s$ being another leading mixture component. 

The left plot of Fig.~\ref{fig:H2nm} depicts that there is an overall 
larger parameter space in the plane of $(\phi_{\kappa}, \phi_{\mu})$ where all the 
physical constraints are satisfied. The requirement of a $H_2$ mass close to $125~\text{GeV}$ 
excludes the region with relatively large $\phi_{\mu}$. As for the BAU allowed regions, we can see it 
is characterized mostly by the horizontal $\phi_{\kappa}$ and is not so sensitive to $\phi_{\mu}$. This is 
because the higgsino-singlino CPV source drives the generation of BAU in this case, i.e., the CPV source given by 
Eq.~\ref{eq:shbau}, and the same for the case of the middle plot of Fig.~\ref{fig:H2nm}.
For the eEDM constraints, the exclusions locate at the lower-left and 
upper-right corners leaving the band between them as viable parameter space. The appearance of this 
viable band between two excluded regions is again due to a cancellation among the three contributing parts (top quark, charginos and $W$ boson loop contributions of the $\gamma H$ Barr-Zee diagrams), which are plotted as contours to help understand the behavior of the eEDM constraints. More explicitly, the magnitude of both the top and W contributions decrease as $\phi_{\mu}$ increases or as $|\phi_{\kappa}|$ decreases. However the $W$ contribution comes with a minus sign and therefore these is a partial cancellation between these two parts. The third part from the chargino contribution increases from a negative value to a large positive value as one goes from the lower-left corner to the upper-right corner. The net effect from all these three contributions gives a negative eEDM at the bottom-left corner of this plot and is excluded by experiment. This negative EDM increases gradually to zero and increases to a larger positive value as $\phi_{\mu}$ increases and as $|\phi_{\kappa}|$ decreases. Therefore a eEDM allowed region is left at the middle of the plot which turns out having relatively large overlap with the BAU and the mass allowed regions.

If we change the signs of $\phi_{\kappa}$ and $\phi_{\mu}$, we then obtain the plot in the middle panel.
In this case, the shape of the excluded region from the mass requirement on $m_{H_2}$ is changed while 
the BAU allowed region shows similar behavior as previous case and is determined mostly by $\phi_{\kappa}$.
For the eEDM constraints, once again, we observe cancellations among the three main contributing parts and 
the behavior of which can be understood with the help of the plotted contours of individual parts.
Putting all these constraints together, we can see there is sizable parameter space in this plane where 
all physical constraints can be satisfied.

The right plot shows the constraints in the plane $(\phi_{\kappa}, \phi_{M_2})$ where we observe no 125 GeV Higgs mass exclusion. As for the eEDM experiment exclusions, it is only the upper-left corner that is excluded
while the majority of the parameter space gives an eEDM prediction that is compatible with the experimental limits. 
This behavior once again results from a cancellation among the three contributing parts. We can see there is a large parameter space that is compatible with all physical constraints in this case. 
Different from the the left and middle plots cases, here the higgsino-singlino CPV source (Eq.~\ref{eq:shbau}) and higgsino-gaugino CPV source (Eq.~\ref{eq:CPVSCH1} and 
Eq.~\ref{eq:CPVSCH2}) together determines the BAU allowed regions.

%\item The case of $H_3$ being SM-like Higgs~\footnote{It should be noted that the $\phi_\mu$ in the two Eqs. should be interpreted as $\phi_\mu-\phi_{M_2}$ since the $\phi_{M_2}$ could be absorbed into the CPV phase of $\mu$ when one study the EDM as in the CPV MSSM scenario~\cite{Lee:2003nta,Bian:2014zka}.} 

\begin{figure}[t]
  \centering
 \includegraphics[width=0.5\textwidth]{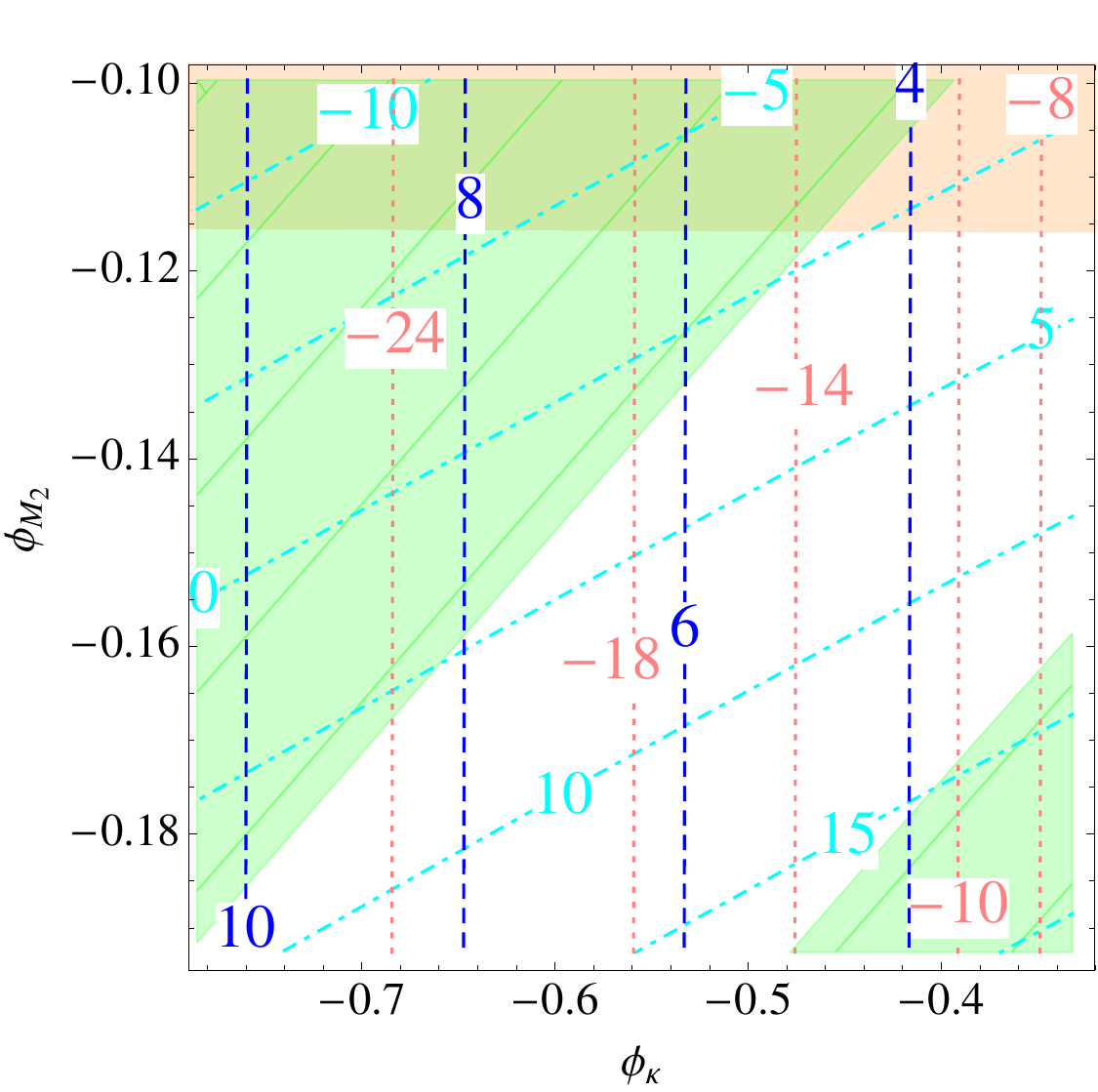}
\caption{
%Fig.6.
\label{fig:H3PQ}
The combined plots of the eEDM and BAU in the plane $(\phi_{\kappa}, \phi_{M_2})$ for the scenario with 
$H_3$ being the SM-like Higgs in the moderate small $\tan\beta=3.05$ of NMSSM.
The color-codes are chosen to be the same as in Fig.~\ref{fig:H2nm}.
Other parameters are set as Table.~\ref{pqbauedm}.
}
\end{figure}

\subsubsection{$H_3$ as the SM-like Higgs for a moderate $\tan\beta$ }

In this section, we study the CPV scenario of NMSSM in the phase plane of
$(\phi_{\kappa}, \phi_{M_2})$ with the $H_3$ being identified as the SM-like Higgs in the approximate PQ limit ($\kappa \approx 0$). In this scenario, $H_1$ is dominated by $h_s$, $H_2$ is dominated by $a_s$, and $H_3$ is dominated by $h_u$ \footnote{ Here, the $H_3$ will not decay to $H_{1,2}$ due to its kinetically forbidden.}. The $H_{1, 2, 3}$ mediated Barr-Zee diagrams dominate the eEDM, all three Higgs are mixture of CP-even Higgs and the CP-odd $a_s$.

\begin{table}[t]
 \begin{center}
 \scalebox{1}{
  \begin{tabular}{c|c|c|c|c|c|c|c|c} \hline
    $\lambda$ &$\mu$ (GeV) & $A_\lambda$ (GeV) & $\kappa$ & $A_\kappa$ (GeV) & tan $\beta$ & $A_{t}(\text{GeV})$\\ \hline
   0.6 & 235 & 838.0 & 0.029 & -99.18 &3.05 & 2024 \\ \hline
$A_{b}(\text{GeV})$&$A_{\tau}(\text{GeV})$&$M_2 (\text{GeV})$ &$M_1(\text{GeV})$&$m_{h_3}$ (GeV) & PTS(CPC)&PTS(CPV)  \\ \hline
1539&1502&200&100&125 & 0.752& 0.833\\ \hline
  \end{tabular}}
 \end{center}
 \caption{The CP conserving NMSSM benchmark point with a moderately small $\tan\beta$, the CPV scenario is obtained by adding imaginary parts for parameters with 
 $\phi_\kappa=0.1$ and $\phi_{M_2}=0.1$. 
\label{pqbauedm}}
\end{table}

In this case, the mass of the SM-like Higgs $m_{H_3}$ is characterized by $\phi_\kappa$ and increases from 126 GeV to 127.5 GeV as $\phi_{\kappa}$ varies from the right side to the left in the region of Fig.~\ref{fig:H3PQ}. Different from the scenarios of the last section, the BAU is determined mostly by $\phi_{M_2}$ and is not very sensitive to the variation of $\phi_{\kappa}$. This is because the BAU is driven mostly by the higgsino-gaugino CPV sources(Eq.~\ref{eq:CPVSCH1} and Eq.~\ref{eq:CPVSCH2}) 
while the contribution from the higgsino-singlino CPV source term (Eq.~\ref{eq:shbau}) is negligible since it is proportional to 
$\phi_{\kappa}$, which is small as a result of the definition of the PQ limit. Here we present a benchmark point in Table.~\ref{pqbauedm} for this moderate $\tan\beta$ scenario. It should be noted that the suppressed weak sphaleron might make the generation of the BAU number harder and the PTS might not be strong enough to prevent the generated BAU being washed out.
As for the eEDM constraints, we can understand the behavior of the eEDM exclusion regions from the contours of individual eEDM contributions. Firstly, the $t/W$ Barr-Zee diagrams contributions are presented by vertical contours in this plots and is characterized solely by $\phi_{\kappa}$. Furthermore, they both give contributions that decreases as the magnitude of $\phi_{\kappa}$ decreases. On the other hand, the charginos contribution to the Barr-Zee diagram of eEDM depends on both CP-phases. Its magnitude decreases first and then increase as 
$|\phi_{M_2}|$ increases or as $|\phi_{\kappa}|$ decreases.

The BAU allowed horizontal band lies at the top of this figure which has 
sizable overlap with the allowed region from EDM constraints. This viable parameter space sits 
at the right-top corner of this plot corresponding to relatively small values of the two phases 
$|\phi_{\kappa}|$ and $|\phi_{M_2}|$.

\section{Direct and Indirect Probe of CPV in the Higgs sector
\label{sec:DP}}
In the CPV NMSSM, there are two types of CPV sources as in aforementioned arguments, i.e., tree level and loop level. The type of the CPV in the Higgs sector is tightly related with the 
CPV in the Higgs mass matrix. When the the CPV occurs at the tree level, the Higgs mass matrix of the complex NMSSM are dominated by the mixing of the SM-like Higgs and heavy CP-odd Higgs like the CPV 2HDM case~\cite{Bian:2014zka,Bian:2016zba}. This type of CPV might be able to be detected through the associated production of Higgs with a ${\bar t} t$ pair~\cite{Gunion:1996vv,Buckley:2015vsa} or a single $t(\bar{t})$~\cite{Ellis:2013yxa}, as well as the Higgs decay into top pairs~\cite{Schmidt:1992et,Bernreuther:1993hq,Bernreuther:2010ny} with top pair polarization being implemented at LHC~\cite{Carena:2016npr} and linear collider~\cite{BhupalDev:2007ftb}, or the $\tau-$lepton decays studies at LHC13~\cite{Harnik:2013aja, Berge:2015nua}. In the meantime,  the CPV in the coupling of $HZ\gamma$ and $HZZ$ can be detected through forward-backward asymmetry of the charged leptons~\cite{Chen:2014ona} and the azimuthal angular distribution of Z boson decay states~\cite{Cao:2009ah}.

%The CPV in the Higgs sector of the complex NMSSM can also induced by the mixing of near degeneracy heavy Higgs mass as in the CPV 2HDM and CPV MSSM scenarios~\cite{Bian:2016zba,Carena:2015uoe}, in this case the CPV phase of loop level dominates over that of tree level, this type CPV arise when the phase combination $\phi^\prime_\lambda-\phi^\prime_\kappa=0$. In this scenario, the CPV would be probed through the analysis of the spins of the $\tau$ and $t$ decay channels at the $\gamma \gamma$ colliders~\cite{Ellis:2004hw} and the LHC~\cite{Ellis:2004fs,Berge:2011ij,Bernreuther:1993df,Berge:2015naf,Chakraborty:2013si,Arbey:2014msa}.  

%In the meantime, the higgsino

\section{Summary and Outlook
\label{sec:Summary}}
With the LHC data accumulating at the 13 TeV energy scale, we would have more access to the Higgs CP properties, which is essential to the cosmic baryon asymmetry generation at the electroweak scale together with a strongly first order phase transition~\cite{Huang:2012wn, Chung:2012vg, Davoudiasl:2012tu}. In this work, we analyse the strength of the EWPT, gravitational wave production and the implementation of EWBG in the CPV NMSSM with large cancellation in the electron EDM allowed by the current eEDM measurements. We first explore the possibility that this model can provide a strongly first order EWPT required to explain the observed baryon asymmetry in the universe at the electroweak scale. Then we investigate the gravitational wave production during the strongly first order EWPT for the benchmark points and show that there exists the possibility for such gravitational waves to be detected in the future space-based interferometers like BBO and Ultimate-DECIGO. We further calculate the baryon asymmetry through the CP violating sources from either higgsino-singlino or higgsino-gaugino and analyze the constraints from the current search limit of the electron EDM. We find that the right amount of baryon asymmetry can be generated without contradicting the EDM limits for small $\tan\beta$ where the 125 GeV SM like Higgs is the second lightest neutral Higgs $H_2$ or moderate $\tan\beta$ where the 125 GeV SM like Higgs is the third lightest neutral Higgs $H_3$ in the CP-violating NMSSM. Such scenarios can be searched in the future for the Higgs CP properties at the LHC or some future neutron/Mercury EDM experiments (footnote 4).  
We note that the BAU evaluation method we adopt in this work(in the Sec.V.A) implicitly assumes the electroweak symmetry being broken inside the bubble and unbroken outside the bubble so that the weak sphaleron process is 
active outside the bubble and quenched inside the bubble.
While, in the two benchmark models being chosen in this work the electroweak symmetry is already broken in the first step of the phase transition (which is a second order phase transition), prior to the second step first order phase transition. Therefore, the weak sphaleron process
outside the bubble might be already exponentially suppressed by a factor of $\exp(-E_{sph}(T)/T)$,
with $E_{sph}$ being the sphaleron energy, and thus highly suppressed the magnitude of the BAU being generated. The successful explanation of BAU requires future comprehensive studies of phase transition in the model with the electroweak symmetry being preserved outside the bubble. Furthermore, one should be aware the cosmologically domain walls problems which is introduced when the $Z_3$ is broken by the singlet of the model.

\section{acknowledgments}
The work of LGB is Supported by the National Natural Science Foundation of China (under
grant No.11605016 and No.11647307), Basic Science Research Program through the National Research Foundation of Korea (NRF) funded by the Ministry of Education, Science and Technology (NRF-2016R1A2B4008759), and Korea Research Fellowship Program through the National Research Foundation of Korea (NRF) funded by the Ministry of Science and ICT (2017H1D3A1A01014046).
The work of JS is supported by the National Natural Science Foundation of China under grant No.11647601, No.11690022 and No.11675243 and also supported by the Strategic Priority Research Program of the Chinese Academy of Sciences under grant
No.XDB21010200 and No.XDB23030100. Part of the results described in this paper are obtained on the HPC Cluster of SKLTP/ITP-CAS.

\section{Appendix: Ingredients for BAU calculations }

 The weak sphaleron rate is $\Gamma_{ws}=6\kappa \alpha_w^5 T$ with $\kappa\approx 20$. 
 In this study, we take wall velocity being $v_w=0.1$ and wall width being $L_W=0.5/T$. The $\Delta\beta\approx 1/M_{H_4}^2$ with $M_{H_4}\sim M_A$ and $M_A$ being the CP-odd Higgs as suggested by Ref.~\cite{Menon:2004wv,Kozaczuk:2013fga}. The diffusion factor are~\cite{Chung:2009qs}: $D_L=100/T$, $D_R=380/T, D_q=6/T$, and $D_H=110/T$.
Related thermal widths are:  $\Gamma_H=0.025 T,~\Gamma_S=0.03 \,T,\Gamma_L=\Gamma_R=0.5 \,T,~ \Gamma_b=0.5 \,T,~\Gamma_{\tilde{b}}=0.5\, T,~\Gamma_{\tilde{\tau}}=0.003\, T,~\Gamma_W=0.065\, T,~\Gamma_B=0.003\, T$.
The $k$ factors  for the calculation of Eq.~\ref{eq:difr} are given by,
$k_{Q,T,B}=k_{q_L,t_R,b_R}+k_{\tilde{q}_L,\tilde{t}_R,\tilde{b}_R}$,~and $k_H=k_{H_d}+k_{H_u}+k_{\tilde{H}}$, with the  k factors for the calculation of Eq.~\ref{eq:difr} in are given by, 
\begin{equation}
k_i(m_i/T) = g_i\frac{6}{\pi^2}\int_{m/T}^\infty dx\,x\,
\frac{e^x}{(e^x \pm 1)^2}\sqrt{x^2 - m^2/T^2}\,,
\end{equation}
 with the $g(H)=g(\tilde{H})=2$, $g(t_L)=g(\tilde{t}_L)=g_{L,B}=3$.

\end{document}